\documentclass[12pt, preprint]{emulateapj}

\usepackage{graphicx}
\usepackage{subfigure}
\usepackage{multirow}	

\usepackage{natbib}
\newcommand{\kms}{~km\,s$^{-1}$}
\newcommand{\ccs}{\,\mathrm{cm}^{-3}\,\mathrm{s}}
\newcommand{\cc}{\,\mathrm{cm}^{-3}}

\newcommand{\vnei}{\texttt{vnei}}
\newcommand{\phabs}{\texttt{phabs}}
\newcommand{\confloop}{\texttt{conf\textunderscore loop}}

\newcommand{\evt}{\texttt{evt2}}
\newcommand{\pha}{\texttt{pha}}
\newcommand{\arf}{\texttt{arf}}
\newcommand{\rmf}{\texttt{rmf}}
\newcommand{\ciao}{CIAO}
\newcommand{\warf}{\texttt{warf}}
\newcommand{\wrmf}{\texttt{wrmf}}
\newcommand{\specextract}{\texttt{specextract}}
\newcommand{\chandra}{\textit{Chandra}}
\newcommand{\XMM}{\textit{XMM-Newton}}
\newcommand{\Laz}{Paper A}

\slugcomment{Published as ApJ, 769, 64}

\shorttitle{The X-Ray Ejecta Knots of Cas A}
\shortauthors{J. Rutherford et al.}

\begin{document}

\title{A Decade-Baseline Study of the Plasma States of Ejecta Knots in Cassiopeia A}

\author{
John Rutherford\altaffilmark{1,2,3}, 
Daniel Dewey\altaffilmark{2,3}, 
Enectali Figueroa-Feliciano\altaffilmark{2,3}, 
Sarah N. T. Heine\altaffilmark{2,3}, 
Fabienne A. Bastien\altaffilmark{4}, 
Kosuke Sato\altaffilmark{5}, and 
C.R.~Canizares\altaffilmark{2,3} }

\affil{Department of Physics and Kavli Institute for Astrophysics and Space Research, \\Massachusetts Institute of Technology, Cambridge, MA 02139, USA; enectali@mit.edu}

\altaffiltext{1}{jmrv@mit.edu} 
\altaffiltext{2}{MIT Kavli Institute, Cambridge, MA 02139, USA}
\altaffiltext{3}{MIT Department of Physics, Cambridge, MA 02139, USA}
\altaffiltext{4}{Department of Physics \& Astronomy, Vanderbilt University, 6301 Stevenson Center, Nashville, TN, 37235}
\altaffiltext{5}{Department of Physics, Tokyo University of Science, 1-3 Kagurazaka, Shinjuku-ku, Tokyo 162-8601, Japan}

\begin{abstract}
We present the analysis of 21 bright X-ray knots in the Cassiopeia A supernova remnant from observations spanning 10 yr. We performed a comprehensive set of measurements to reveal the kinematic and thermal state of the plasma in each knot, using a combined analysis of two high energy resolution High Energy Transmission Grating (HETG) and four medium energy resolution Advanced CCD Imaging Spectrometer (ACIS) sets of spectra. The ACIS electron temperature estimates agree with the HETG-derived values for approximately half of the knots studied, yielding one of the first comparisons between high resolution temperature estimates and ACIS-derived temperatures. We did not observe the expected spectral evolution—predicted from the ionization age and density estimates for each knot—in all but three of the knots studied. The incompatibility of these measurements with our assumptions has led us to propose a dissociated ejecta model, with the metals unmixed inside the knots, which could place strong constraints on supernova mixing models.
\end{abstract}

\keywords{
supernovae: individual (Cas A) --
techniques: imaging spectroscopy --
X-rays: ISM --
ISM: supernova remnants --
plasmas -- 
shock waves
}

\section{Introduction}

Supernova remnants (SNRs) are unique laboratories for studying astrophysical plasmas, nuclear physics, and chemical evolution. The hot, low-density plasma formed in the wake of the supernova explosion shock wave cannot be created in laboratories, or even in the coronae of nearby stars. The ejecta from the supernova provide the metals for the enrichment and chemical evolution of the galaxy \citep{Matteucci:1986ux, Pagel97,Fukugita:2004bf}.

The SNR Cassiopeia A (Cas A), distinguished by its filamentary bright X-ray features, has been well studied in the X-ray band. From the \chandra\ first light observation \citep{Hughes00}, these features have been identified as Si-rich ejecta knots, coherent material expelled from the deeper layers of the progenitor star. As \cite{LH03} first reasoned, the term ``knot'' here refers to the bright X-ray features having densities of tens to hundreds of electrons per cubic centimeter --- which is only a factor of a few above the density of the surrounding material --- in contrast to the much denser ``optical knots'' of \cite{Fesen11}. \citeauthor{LH03}, along with the companion paper \cite{HL03}, systematically investigated the plasma properties of X-ray knots first, using the regions as tracers to infer the surrounding envelope's density profile. \citet{Lazendic} (\Laz\ hereafter) performed the first high spatial-spectral resolution analysis of the bright X-ray knots, deriving temperatures, Doppler velocities, and abundances. That analysis utilized dispersed spectra of the extended remnant, a difficult and uncommon technique that our current work also employs. The entire 3-dimensional structure of Cas A was fleshed out beyond the skeleton of these bright knots by \citet{DeLaney10}. The exhaustive investigation suggests a flattened explosion and ejecta ``pistons''. Recently, \cite{HL12} finely mapped abundances and plasma states across the entire remnant for the 1 Ms \chandra\ observation in an analysis tour de force.

These thermal X-ray knots are the subject of our investigation. We sought to characterize the plasma state of a set of localized ejecta regions in Cas A and investigate their evolution over a decade in the remnant's approximately 330 year lifetime. We collected both imaging and dispersed data of Cas A with the Advanced CCD Imaging Spectrometer (ACIS) and High Energy Transmission Grating Spectrometer (HETGS) instruments, respectively, on board \chandra. Analysis of the dispersed data yields ratios of the strongest lines (we only look at Si), but these ratios alone cannot fully describe the plasma state, including the electron temperature and ionization age. We therefore use the broadband spectra from the imaging data --- with a model --- to provide a fuller picture of the plasma state. We address two primary questions regarding the physical state of these knots. First, over the decade of observations of Cas A, can we see these knots evolving spectrally? Second, how do the two pictures of the knots' plasma states --- the broadband plasma model parameters from the ACIS analysis and the Si line ratios from the HETGS data --- compare and agree with predictions?

\begin{deluxetable*}{cccccc}
\tablewidth{0pt}
\tablecaption{Summary of HETG observation parameters}
\tablehead{ ObsID & Start Date & Exposure (ks) & RA & DEC & Roll Angle }
\startdata
1046	 & 2001 May 25 & 69.93 & $23^{\mbox{h}} 23^{\mbox{m}} 28^{\mbox{s}}.00$ & $+58^{\circ} 48^{\prime} 42^{\prime\prime}.50$ &  85.96 \\
10703 & 2010 Apr 27 & 35.11 & $23^{\mbox{h}} 23^{\mbox{m}} 27^{\mbox{s}}.90$ & $+58^{\circ} 48^{\prime} 42^{\prime\prime}.50$ & 60.16 \\
12206 & 2010 May 2 & 35.05 & $23^{\mbox{h}} 23^{\mbox{m}} 27^{\mbox{s}}.90$ & $+58^{\circ} 48^{\prime} 42^{\prime\prime}.50$ & 60.16 
\enddata
\label{tab:HETGobs}
\end{deluxetable*}

Cas A has been the subject of several other long-term investigations recently. \citet{Patnaude07} took a similar approach to this work and looked at the evolution of four bright regions over four years, observing flux variation and some changes in plasma parameters. A full decade's worth of high quality \chandra\ data on Cas A was utilized by \citet{Patnaude11}. That study found a 1.5\% per year decrease in the nonthermal X-ray flux across the remnant, while the thermal component stayed fairly constant. In the optical band, \citet{Fesen11} studied the flickering phenomenon of outer ejecta knots over half a decade, which can reveal the initial enrichment of the interstellar medium. 

Due to the low densities of these knots and relatively short timescales, the ions and electrons have not interacted enough to achieve collisional ionization equilibrium (CIE), remaining instead in non-equilibrium ionization (NEI). Ionization age is a measure of the evolution of a plasma from NEI toward CIE; $\tau = n_{e} t$, where $n_{e}$ is the electron density in $\cc$ and $t$ the time in seconds since encountering the reverse shock. As $\tau$ increases over time, the electron temperature increases to meet the the proton and ion temperatures. In NEI the fractions of the different ionization species for a given element evolve with time as the ionization and electron capture processes balance. The CIE timescale depends on the plasma temperature and its elemental abundances; for a Si-dominated plasma at 2~keV the timescale is $\tau \sim 10^{12}\ccs$~\citep{Smith:2010bs}. The ionization ages of around $10^{10}\ccs$ derived in \Laz\ for several knots suggested the knots were in various states of NEI. The derived electron densities $n_{e}\sim 100\cc$ implied that significant signs of plasma evolution should be evident over a ten-year period, through a measurable increase of the ACIS-derived ionization age and an increased Si XIV to Si XIII emission ratio in the HETG data.

We report results from analyses of observations over a ten year baseline, both with a new HETG observation (Table~\ref{tab:HETGobs}), and existing ACIS data (Table~\ref{tab:ACISobs}). We focus on 21 knots, which include the 17 knots studied in \Laz. A comprehensive set of measurements of the thermal and kinematic properties of these knots is presented. Unlike \Laz, wherein plasma parameters like temperature were inferred from line ratios, our analysis directly models and fits these parameters with ACIS data. We thereby obtain independent high-resolution line ratios and plasma parameters, ensuring cross-validation for our results.

This study finds significantly higher temperature estimates and lower ionization ages than \Laz\ for most knots. The rapid evolution predicted from these new values was largely not seen. These data present strong evidence that one cannot model these knots as single-temperature non-equilibrium systems evolving without interactions with their surroundings. We propose a dissociated model of the metals inside the knots, which could constrain supernova explosion models. 

The observations and analysis are described in Section~\ref{sec:obs}, with the results presented in Section~\ref{sec:results} and discussed in Section~\ref{sec:disc}. The details of the analysis are presented in Appendices~\ref{sec:HETGan} (for HETG) and \ref{sec:ACIS-anal} (for ACIS). The tabulated results and several plots for all the knots are presented in Appendix ~\ref{sec:result-app}.

\section{Observations and Analysis}
\label{sec:obs}

\begin{figure}[t]
\begin{center}
\includegraphics[angle=0,width=0.95\columnwidth]{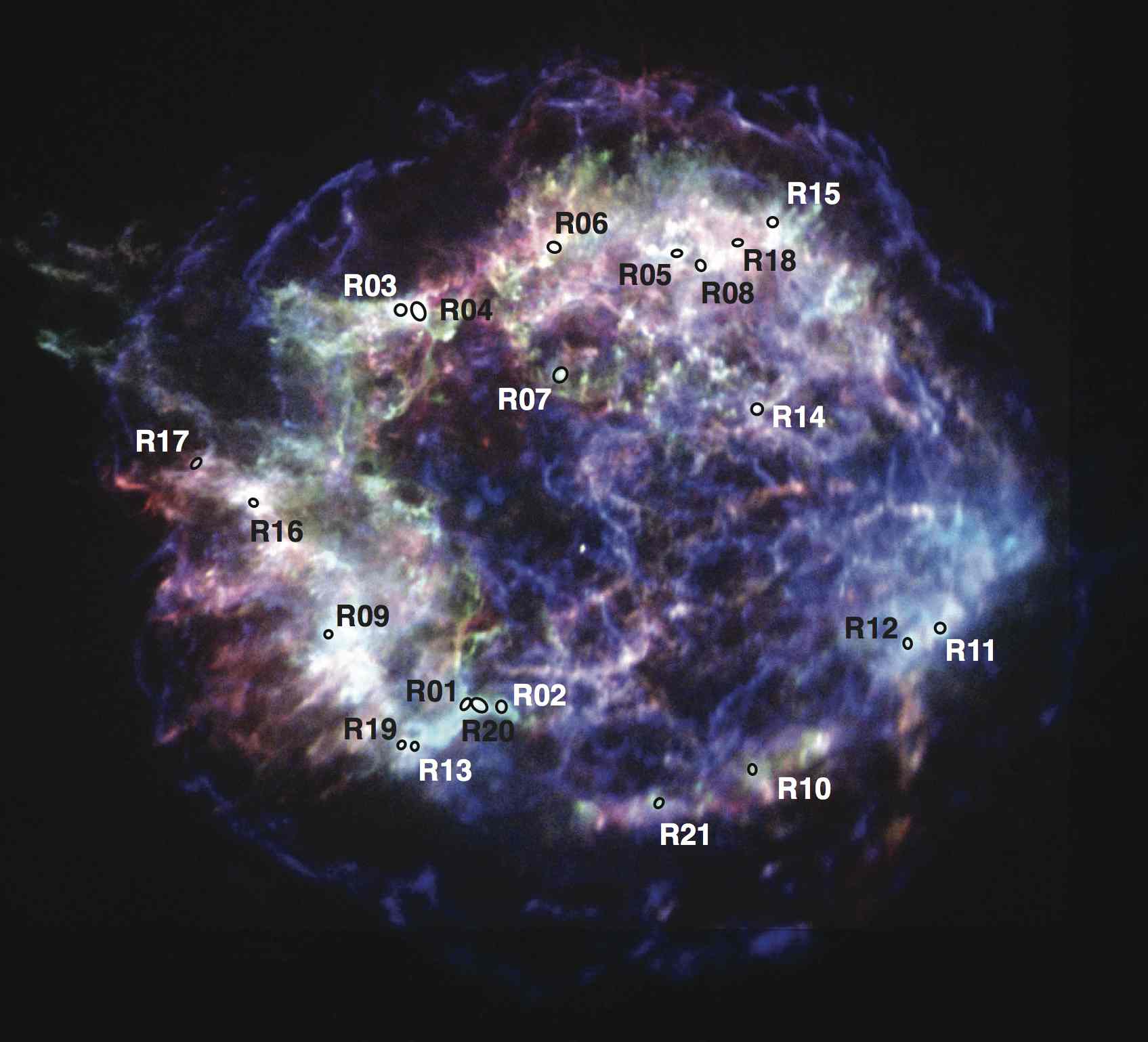} \\			
\end{center}
\caption{We investigated 21 bright Si X-ray knots in Cas A for this study. The first 17 coincide with those in  \cite{Lazendic}. R18 and R21 have been added, as they have brightened, and R19 and R20 were split off of R13 and R01, as they have separated on the sky since the first observation.
\label{fig:knotLocs}}
\end{figure}

We chose 21 knots for our analysis. These are shown in Figure~\ref{fig:knotLocs} and listed in Table~\ref{tab:knotLocsTable}. The first 17 were defined to coincide with those in \Laz, which were selected for their bright Si features. Four more were added: R18 and R21 because they have brightened; R19 and R20 were split off of R13 and R01 because they have separated on the sky since the first observation.

The data we collected for this work are of two types: gratings-dispersed images, which have higher spectral resolution, and non-dispersed images, with moderate spectral resolution. For the dispersed HETG analysis, a new Monte Carlo method called Event-2D was used to extract the Si lines of the knots from the very complex dispersed datasets. For the non-dispersed ACIS analysis, we compared the derived plasma model parameters of these knots over the decade. The fine-grained spectral detail of the HETG work and the broad-brush nature of the ACIS analysis complement each other to provide a better picture of the physical processes in Cas A.

\subsection{High Spectral Resolution: HETG}

\begin{deluxetable}{ccc}
\tablewidth{0pt}
\tablecaption{Summary of ACIS observation parameters}
\tablehead{ ObsID & Start Date & Exposure (ks) }
\startdata
114 & 2000 Jan 30 & 50.56 \\
4637 & 2004 Apr 22 & 165.66 \\
9117 & 2007 Dec 5 & 25.18 \\
10935 & 2009 Nov 2 & 23.58 
\enddata
\label{tab:ACISobs}
\end{deluxetable}

\begin{figure}[t]
    \begin{center}
        \subfigure[]{
            \includegraphics[width=\columnwidth]{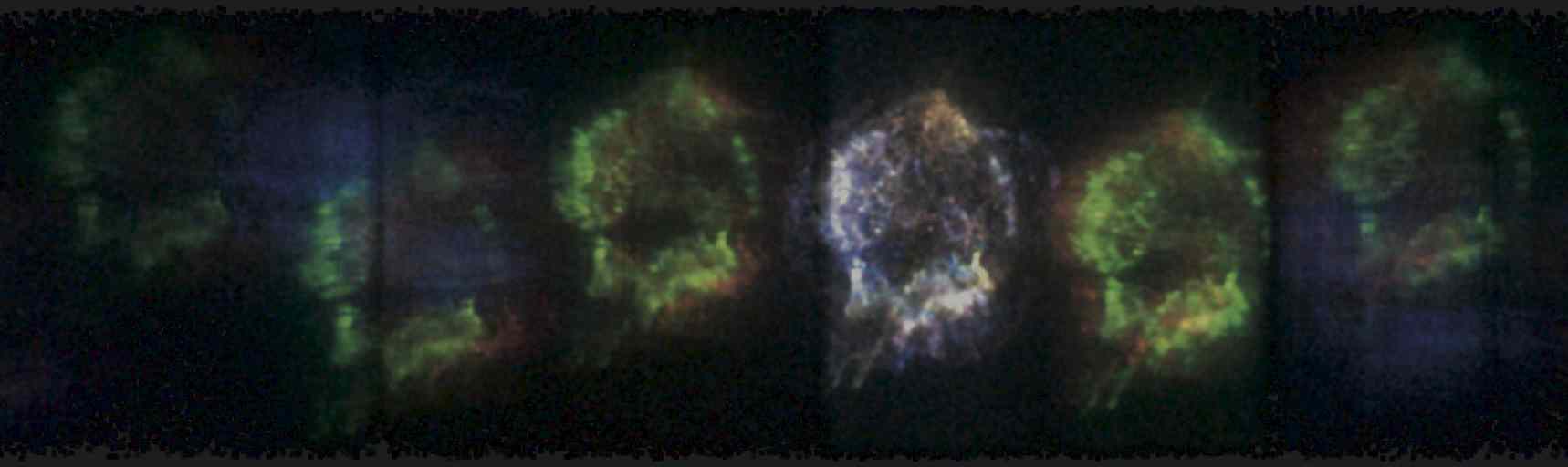}
        } \\
        \subfigure[]{
           \includegraphics[width=\columnwidth]{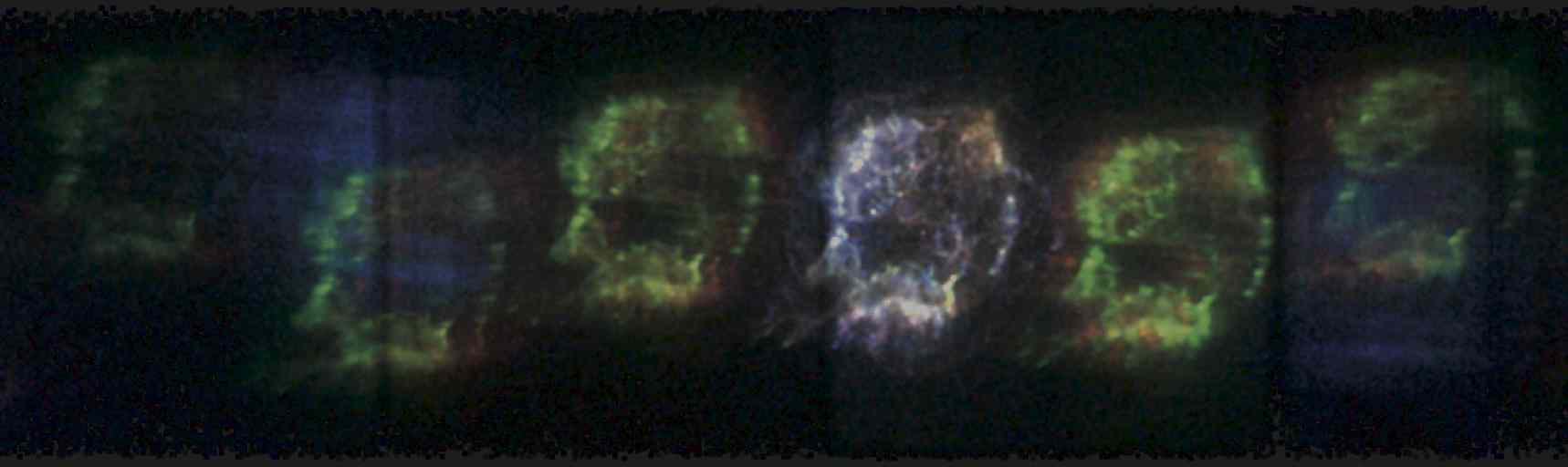}
        }
    \end{center}
    \caption{The bright knots of Cas A can be analyzed using dispersive gratings. The two HETG observations of Cas A --- 2001.4 in (a), 2010.3 in (b) --- were taken at different roll angles. Left to right, the images show orders MEG-3, HEG-1, MEG-1, zeroth, MEG+1, HEG+1. The false color image shows the Si XIV line (red), the Si XIII blend (green), and lines below 1 keV (blue). \\
    \label{fig:HETGobs}}
\end{figure}

The prospect of directly measuring temperature or ionization state evolution for X-ray knots -- based on the results of \Laz\ in 2001-- motivated a second HETG observation of Cas A in 2010. The observations span a nine-year baseline, and were taken as part of the \chandra\ GTO program (Table~\ref{tab:HETGobs}). The latter observations were taken at a different roll angle than the \Laz\ dataset, affording the ability to cross-check the data reduction process, as the knots would be dispersed across different slices of Cas A. Figure~\ref{fig:HETGobs} shows how the bright features of Cas~A can be tracked over multiple dispersive orders.

For the HETG analysis we used --- for the first time --- the Event-2D technique \citep{Dewey09}. This complex high-energy-resolution analysis incorporates a 2D model of each knot and its background, and folds it through the response function of the HETG, to generate a model image that includes the 2D morphology, the spectral model, and the dispersion response. This model image is then compared to the actual data image collected with the ACIS-S CCD, and fit with a Monte Carlo technique. The method yields increased energy resolution over naive approaches by modeling complex shapes and utilizing the CCD spectral resolution to disentangle the source from the background. 

The Event-2D analysis uses a 4-Gaussian model corresponding to the Si~XIII (He-like) recombination, inter-combination and forbidden ($r$, $i$ and $f$, respectively) lines and the Si~XIV (H-like) Ly\,$\alpha$ line.  The $f/i$ ratio is fixed at the low density value of 2.45 and the remaining free parameters are then the overall flux, the $f/r$ ratio, the H-like/He-like ratio and the Doppler velocity. A thorough description of this analysis is detailed in Appendix~\ref{sec:HETGan}.

\subsection{High Spatial Resolution: ACIS}

To complement the new HETG data, four archival ACIS observations spanning a ten-year baseline were analyzed (Table~\ref{tab:ACISobs}), from early 2000, mid-2004, late 2007, and late 2009. 

We assume these small, bright knots are individual clumps of similar material, so we therefore defined their boundaries spectrally. Arbitrarily bounding these regions by hand, using only brightness information, could unintentionally adulterate the knot's spectrum with non-knot material of similar surface brightness, thereby skewing results of the analysis. Our automated, systematic method found where the surrounding spectrum differed significantly from that of the central knot core, and drew the knot boundary accordingly.

We used the fainter regions surrounding the knots as the ``background,'' choosing diffuse plasma showing little variation with position, under the assumption that the spectral core we extracted from the knot region contained this diffuse material either in front of or behind it.

With the regions and backgrounds defined, we extracted the data using a custom pipeline based on the \specextract\ \ciao\ script that correctly takes into account the dithering of the telescope during the observation, which is important at the small scales of our knot regions. 

\begin{deluxetable}{ccc}[h]
\tablewidth{0pt}
\tablecaption{The coordinates of the knot centers in the 2004.4 epoch.
\label{tab:knotLocsTable}}
\tablehead{ Region & RA & Dec}
\startdata
R01&$23^{\mbox{h}}23^{\mbox{m}}33^{\mbox{s}}.176$&$+58^{\circ}47^{\prime}47^{\prime\prime}.34$\\
R02&$23^{\mbox{h}}23^{\mbox{m}}31^{\mbox{s}}.566$&$+58^{\circ}47^{\prime}46^{\prime\prime}.73$\\
R03&$23^{\mbox{h}}23^{\mbox{m}}35^{\mbox{s}}.934$&$+58^{\circ}50^{\prime}03^{\prime\prime}.80$\\
R04&$23^{\mbox{h}}23^{\mbox{m}}35^{\mbox{s}}.199$&$+58^{\circ}50^{\prime}04^{\prime\prime}.59$\\
R05&$23^{\mbox{h}}23^{\mbox{m}}23^{\mbox{s}}.626$&$+58^{\circ}50^{\prime}23^{\prime\prime}.88$\\
R06&$23^{\mbox{h}}23^{\mbox{m}}29^{\mbox{s}}.187$&$+58^{\circ}50^{\prime}26^{\prime\prime}.64$\\
R07&$23^{\mbox{h}}23^{\mbox{m}}28^{\mbox{s}}.843$&$+58^{\circ}49^{\prime}43^{\prime\prime}.34$\\
R08&$23^{\mbox{h}}23^{\mbox{m}}22^{\mbox{s}}.554$&$+58^{\circ}50^{\prime}20^{\prime\prime}.83$\\
R09&$23^{\mbox{h}}23^{\mbox{m}}39^{\mbox{s}}.236$&$+58^{\circ}48^{\prime}12^{\prime\prime}.36$\\
R10&$23^{\mbox{h}}23^{\mbox{m}}20^{\mbox{s}}.273$&$+58^{\circ}47^{\prime}25^{\prime\prime}.01$\\
R11&$23^{\mbox{h}}23^{\mbox{m}}11^{\mbox{s}}.871$&$+58^{\circ}48^{\prime}14^{\prime\prime}.00$\\
R12&$23^{\mbox{h}}23^{\mbox{m}}13^{\mbox{s}}.388$&$+58^{\circ}48^{\prime}09^{\prime\prime}.34$\\
R13&$23^{\mbox{h}}23^{\mbox{m}}35^{\mbox{s}}.506$&$+58^{\circ}47^{\prime}33^{\prime\prime}.70$\\
R14&$23^{\mbox{h}}23^{\mbox{m}}20^{\mbox{s}}.101$&$+58^{\circ}49^{\prime}31^{\prime\prime}.14$\\
R15&$23^{\mbox{h}}23^{\mbox{m}}19^{\mbox{s}}.376$&$+58^{\circ}50^{\prime}35^{\prime\prime}.02$\\
R16&$23^{\mbox{h}}23^{\mbox{m}}42^{\mbox{s}}.746$&$+58^{\circ}48^{\prime}58^{\prime\prime}.16$\\
R17&$23^{\mbox{h}}23^{\mbox{m}}45^{\mbox{s}}.045$&$+58^{\circ}49^{\prime}12^{\prime\prime}.76$\\
R18&$23^{\mbox{h}}23^{\mbox{m}}20^{\mbox{s}}.914$&$+58^{\circ}50^{\prime}28^{\prime\prime}.55$\\
R19&$23^{\mbox{h}}23^{\mbox{m}}35^{\mbox{s}}.935$&$+58^{\circ}47^{\prime}34^{\prime\prime}.50$\\
R20&$23^{\mbox{h}}23^{\mbox{m}}32^{\mbox{s}}.519$&$+58^{\circ}47^{\prime}47^{\prime\prime}.61$\\
R21&$23^{\mbox{h}}23^{\mbox{m}}24^{\mbox{s}}.498$&$+58^{\circ}47^{\prime}13^{\prime\prime}.35$
\enddata
\end{deluxetable}

\begin{figure}
\begin{center}
\includegraphics[angle=0,width=\columnwidth]{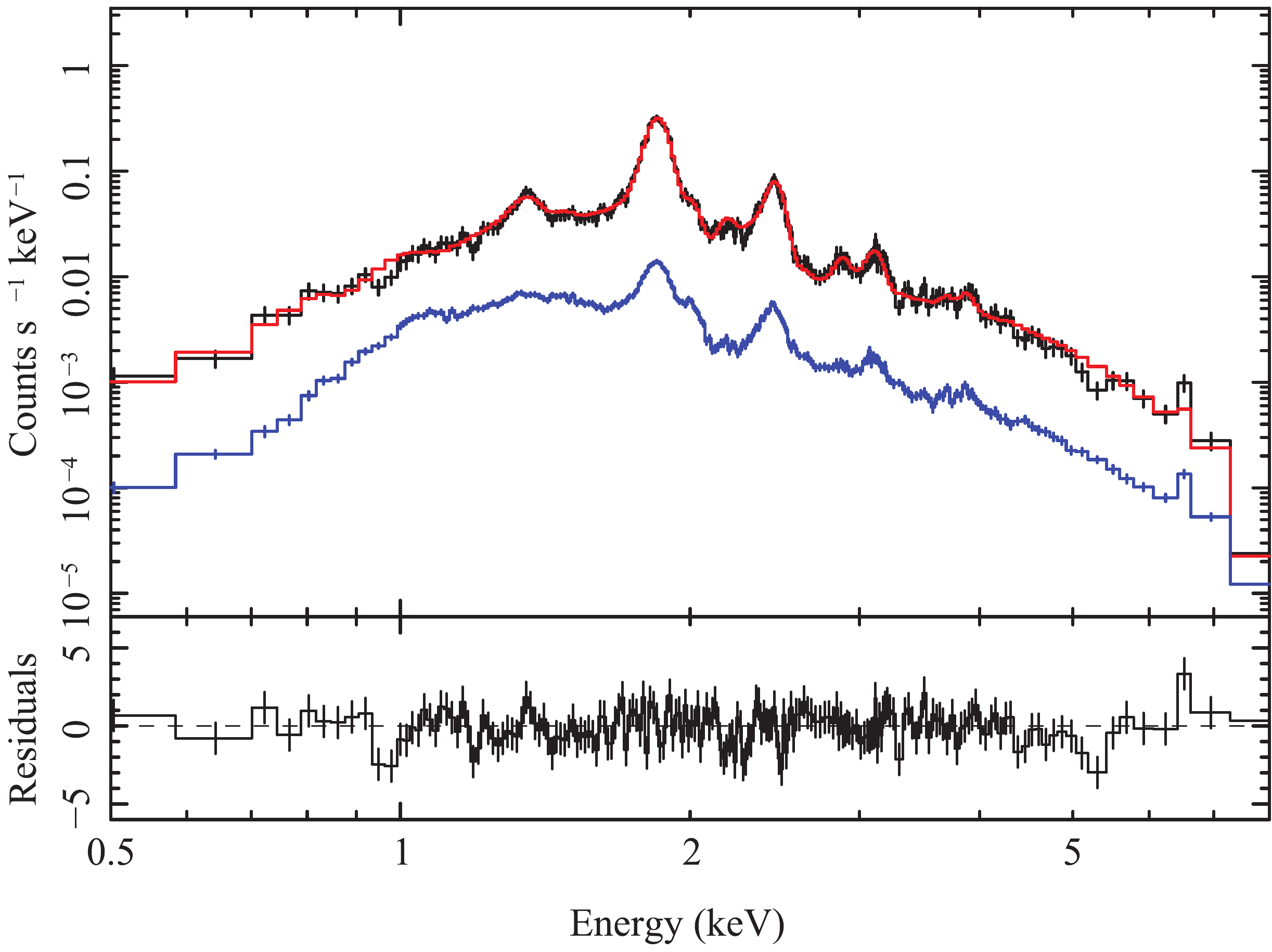} 
\end{center}
\caption{The \vnei\ model fits the ACIS data well; shown is the 2004.4 epoch for R12. The data appear in black, the background in blue, and the combined model in red. \\
\label{fig:acisspectrum}}
\end{figure}

Each knot was fit with the conventional model: a non-equilibrium ionization (NEI) plasma with variable abundances (\vnei, \citet{borkowski01}), with the equivalent absorption by neutral hydrogen ($n_{H}$) between the source and observer accounted for (\phabs, abundances from \citet{bcmc}). Figure~\ref{fig:acisspectrum} shows a typical fit to the data for a single epoch. All fitting was performed with the ISIS software\,\citep{Houck00}. As we assume the foreground and knot compositions do not change over our observation period, we performed fits to the four epochs of ACIS data for each knot simultaneously, tying $n_{H}$ and the abundances across all datasets.  The redshift was fixed after fitting it to the bright Si lines. (For our radial velocity measurements, we used the more accurate HETG Doppler shift.) The final values of the tied parameters were dominated by the 2004.4 dataset due to its longer integration time. The overall scaling factor (the norm), the electron temperature $kT$, and ionization age $\tau$ were allowed to float independently for each epoch in the combined fit. Pileup corrections were also used. The details of the ACIS analysis are presented in Appendix~\ref{sec:ACIS-anal}.

Counter to other analyses (\Laz, \citeauthor{HL12}), we include the contributions of low-Z elements other than oxygen to the spectral continuum. H and He are held at solar values, and C, N, and O are fixed at 5 times solar. As will be discussed in Section~\ref{sec:vneidegen}, the model and data cannot distinguish between which elements prop up the continuum, so these values are merely fiducial. We include $Z<8$ for two reasons: (1) nucleosynthesis models predict yields of C and N on order with O, and (2) \citet{Dewey:2007ue} argue that O cannot account for all of the continuum in Cas A, otherwise the line emission would be too strong.

\section{Results}
\label{sec:results}

The two analysis pipelines produced several independent plasma diagnostics: primarily, the f/r ratio in Si XIII, the Si XIII/XIV line ratio, the ionization age from \vnei, and the electron temperature, also from the \vnei model. For considerations of flow and space, we present the full tabular results of our analyses in appendices.

The HETG analysis was performed for the observations in Table~\ref{tab:HETGobs}, and the results of those fits are shown in Appendix~\ref{sec:result-app}, Figure~\ref{fig:hetgvalues} and Table~\ref{tab:HETGresults}. Uncertainties are given for all but the flux, which has a statistical uncertainty generally less than 2\% and so will be dominated by systematic errors, such as the calibration of the effective area at a knot's specific location.

The ACIS analysis was performed for the observations in Table~\ref{tab:ACISobs}, and the results of these fits are shown in Appendix~\ref{sec:result-app}, Figures~\ref{fig:acisresults1} -- \ref{fig:acisresults3}, Tables~\ref{tab:ACISresults}, and \ref{tab:acistied}. The reported parameter ranges represent bounding edges of confidence contours (see Figure \ref{fig:ccedges} and Appendix~\ref{sec:ACIS-anal}).

\subsection{Comparison with \Laz\ \\ and other results}

The values of the HETG line ratios from the first epoch are consistent with those obtained in the \Laz\ analysis, validating the Event-2D technique. The latest HETG observation and new multi-epoch fits to the ACIS data extend their work, and challenge their conclusions. 

Our temperature and ionization age results differ substantially from \Laz. \Laz\ used Si line ratios to infer $kT$ and $\tau$ values by comparing to the XSPEC model ratios. We believe our ACIS analysis of four combined epochs has much smaller systematics to directly determine $kT$ and $\tau$. Perhaps this tension is not too surprising, though; \Laz\ derived the parameters from Si only, while we infer $kT$ and $\tau$ from a broadband spectrum composed of several ion species and a continuum component. More specifically, while all but three --- R08, R10, and R17 --- of \Laz 's knot temperatures are \emph{lower} than $kT=1.6$~keV and are clustered around $kT\simeq 1$~keV, all of our results are \emph{higher} than $kT=1.6$~keV and are clustered around $kT\simeq 2.2$~keV (see Figures~\ref{fig:acisresults1} -- \ref{fig:acisresults3}). With respect to ionization age, Table~\ref{tab:ACISresults} lists systematically lower $\tau$ values than \Laz . All but two of our knots --- R09 and R16 --- have $\tau$ lower than $10^{11}\ccs$, while all but four --- R07, R08, R10, and R17 --- of \Laz 's values are above $10^{11}\ccs$. 

This work also portrays a slightly different ionization state than investigations by \citeauthor{HL12}. Their mosaic of Cas A for the 2004.4 epoch gave distributions of $kT$, peaked at 1.4 keV, and $\tau$, peaked at $2.5 \times 10^{11}\ccs$ \citep{HL12}. This handful of knots, however, grouped around $kT = 2.2$ keV and $\tau = 4 \times 10^{10}\ccs$. The discrepancy could be due to choice of model (\texttt{vpshock} vs. \texttt{vnei}), or simply show these knots to be a different population. The Si and S abundances are higher than in the mosaic regions, which is to be expected, but the other elements are consistent. \citet{LH03} analyzed several of the same knots: R03 $=$ NE8, R04 $=$ NE6 $+$ NE7, R08 $=$ NNW4. We see general agreement, except for R08, which exhibits double the temperature and half the ionization age. The Si/O ratios are consistently twice as high in this analysis, though we fix the O abundance. None of these discrepancies are cause for concern, especially given the different analysis schemes.

The abundances of Si and S are consistent with nucleosynthesis models, though cannot constrain them. Even though these knots are local, Si-rich features, it is instructive to compare the prominent elements with fiducial global abundance predictions. We turn the abundance point estimates from Table~\ref{tab:acistied} into a mass ratio with the solar abundances from \citet{ag89}, resulting in a large spread: $1.4 \leq M_{Si}/M_S \leq 2.4$. \citet{rauscher} consider several masses for a Type II progenitor, and two sets of reaction rates, resulting in mass ratios $1.9 \leq M_{Si}/M_S \leq 2.5$. 

\bigskip
\bigskip
\bigskip

\subsection{Theory approaching observation: \\ two measurements of temperature}

This collection of knots forms a unique set of isolated low density plasmas, the characteristics of which can be compared against the theoretical predictions of atomic transition codes. Table~\ref{tab:HETGresults} shows the He-like $f/r$ and H/He ratios of Si for this sample of knots from the HETG analysis. The 90\% confidence bounding edges for $kT$ and $\tau$ are shown in Table~\ref{tab:ACISresults}.

\begin{figure}
\begin{center}
\includegraphics[angle=0,width=\columnwidth]{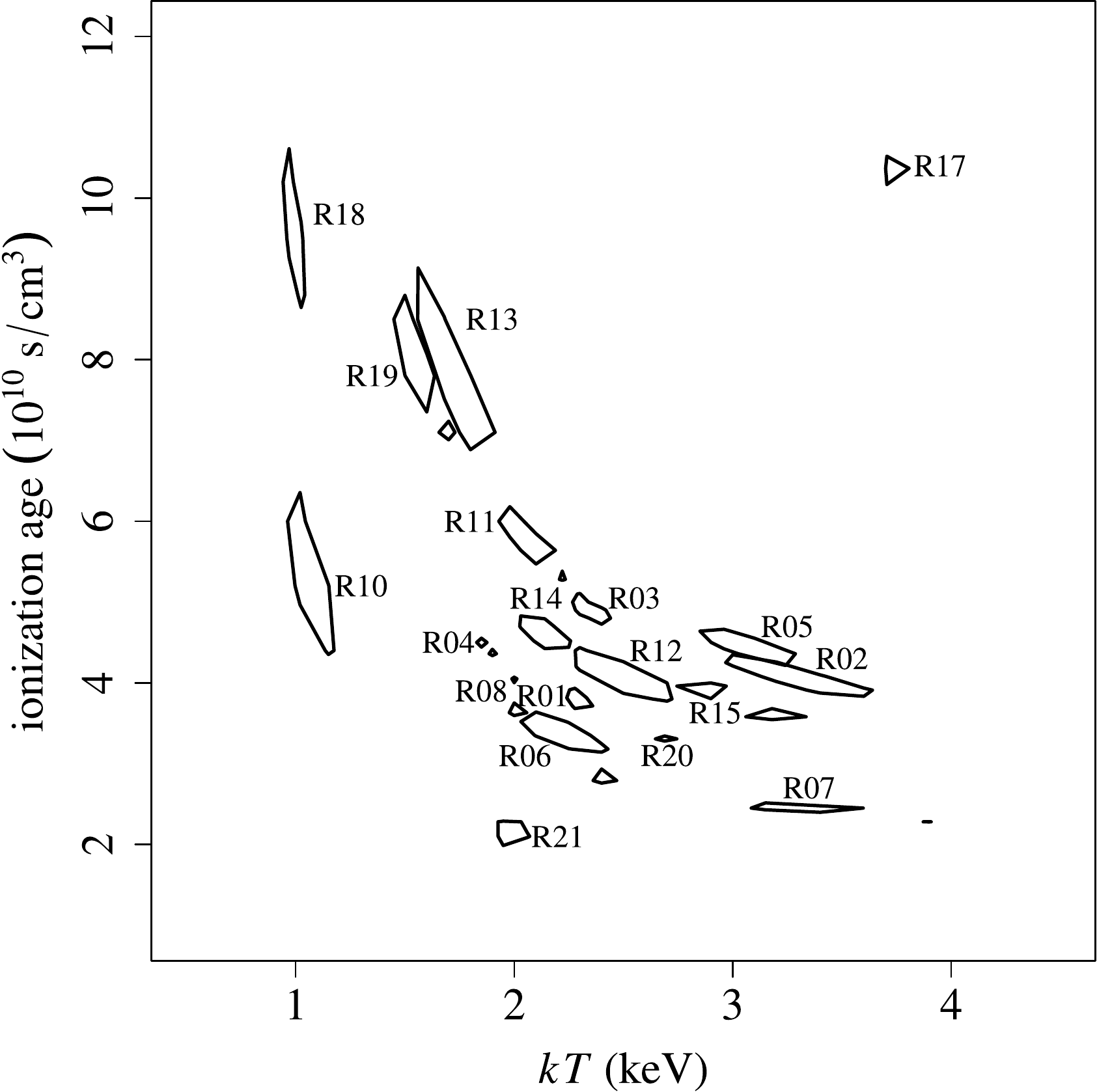} 
\end{center}
\caption{
The knots exhibit a variety of plasma states. 68\% confidence contours in the $kT$--$\tau$ plane for knots R01--R20 in the 2004.4 epoch display systematically larger values than \Laz . A large anti-correlation between $kT$ and $\tau$ can be seen. This correlation widens the confidence intervals, compared to the 1D confidence interval obtained from ISIS or XSPEC, and it highlights the pitfall of reporting 1D confidence levels without accounting for correlations in the fit parameters.
\label{fig:1sigma}}
\end{figure}

The knots exhibit a variety of ionization states. The Si XIII $f/r$ ratios run from 0.18 to to 0.88, while the Si H/He ratios similarly cover a wide range, from to 0.035 to 0.60. The ACIS analysis revealed knots covering much of $kT$--$\tau$ space, with temperatures from below 1 to above 7 keV, and ionization ages from 2 to above $50 \times 10^{10}\ccs$, as can be seen in Figure~\ref{fig:1sigma}.

He-like transition ratios have long been used as temperature diagnostics. \citet{Gabriel69} applied the technique to derive temperatures in the corona of the Sun, where electron densities reach $10^{11}\mbox{ cm}^{-3}$. With an eye toward extrasolar \chandra\ observations, \citet{Porquet01} significantly improved calculations for ions out to Si XIII in collisional plasmas. Most recently, \citet{Smith09} applied new fully relativistic code to Ne IX, demonstrating significant corrections to previous calculations. (More complete references on the utility of He-like transitions can be found therein.)

The HETG and ACIS analyses yield two different measurements of temperature, which we may compare against each other. The ratios of Si XIII lines from the HETG analysis can be modeled with atomic transition codes to infer the temperature of the emitting plasma, while the ACIS analysis produces a temperature which is interpreted within the \vnei\ model. We note that emission models in such atomic transition codes assume CIE, while our ACIS model assumes NEI. 

The ionization age, $\tau$, provides a measure of the stage of evolution of the plasma from NEI to CIE. Since at large ionization age an NEI plasma will come to equilibrium and asymptote to CIE~\citep{Smith:2010bs}, we would expect good correlation between the line-ratio-derived temperature and the \vnei -derived temperature for plasmas with large $\tau$. 

\begin{figure}
\begin{center}
\includegraphics[angle=0,width=\columnwidth]{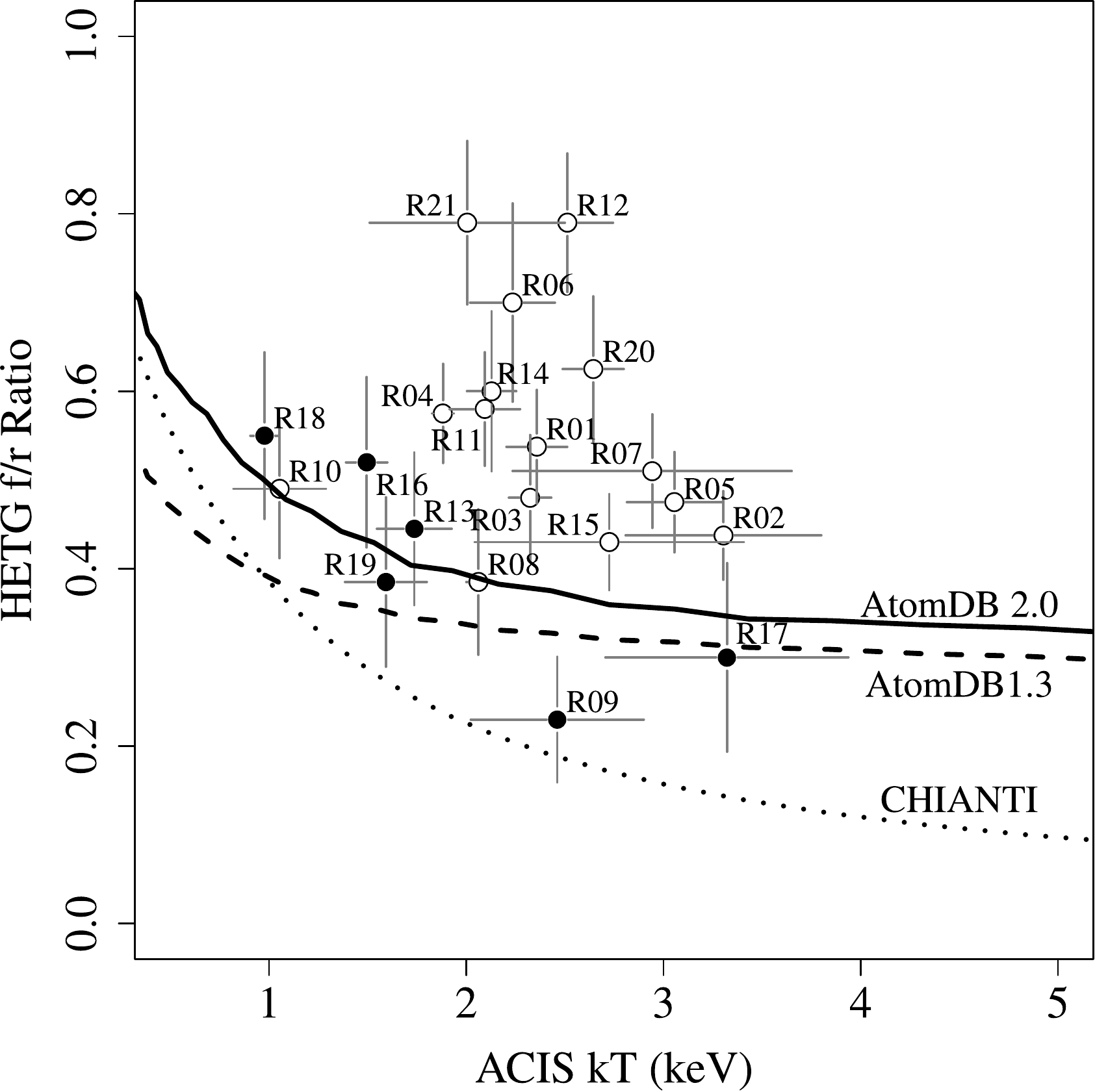} \\
\end{center}
\caption{
Modern plasma emission codes (curves) cannot fully describe the ACIS-derived temperatures ($x$-axis) and HETG-derived f/r ratios ($y$-axis) of the 21 ejecta knots. The ``NEI outliers'' (white dots), which differ most significantly from the plasma emission codes, have systematically lower ionization ages ($\tau < 6 \times 10^{10}\mbox{ s cm}^{-3}$). These knots therefore break the collisional ionization equilibrium assumption implicit in the plasma database calculations. 68\% confidence and 1-$\sigma$ error bars are shown for the knots  along the $x$- and $y$-axes, respectively. The knots are identified by the knot number from Figure~\ref{fig:knotLocs}. The \chandra\ measurements are compared against the CHIANTI 6.0 database (dotted), AtomDB 1.3.1 (dashed), and AtomDB 2.0.0 (solid). 
\label{fig:apedComparison}}
\end{figure}

Figure \ref{fig:apedComparison} shows a comparison of the emission line-derived and model-derived temperatures. The $y$-axis shows the average $f/r$ value of the two HETG observations, and the $x$-axis shows the temperature for the 2004.4 ACIS observation. As most knots did not show much evolution, the average and the mid-decade values are appropriate. 

Nearly half of the knots are consistent with the latest plasma emission code, AtomDB 2.0 \citep{atomdb2}. Two other plasma emission codes, an earlier version of AtomDB and the latest CHIANTI database \citep{chianti2}, predict lower $f/r$ ratios over the same temperature range.  All databases are evaluated in the low density limit. The new calculations of excitation and recombination rates have brought the new AtomDB into greater agreement with these knot observations.  

The ``NEI outliers'' above the curve exhibit too much forbidden Si XIII line radiation for their derived temperatures, by a factor of two in some cases. These outliers exhibit low, tightly grouped ionization ages. Figure \ref{fig:apedComparison} shows the two different distributions for the 2004.4 epoch maximum likelihood values. The outlier distribution is clustered around an ionization age of $\tau = 5\times 10^{10}\ccs$, indicating a plasma fairly far from CIE~\citep{Smith:2010bs}. These knots have been caught early in the evolution process, so they deviate most from the AtomDB ionization equilibrium calculations. The knots closer to the AtomDB 2.0 line vary more widely in derived $\tau$ values, though some show equally low values as the NEI outliers. The populations do not differentiate as clearly in \emph{kT}, H/He, or $f/r$. 

The results in Figure \ref{fig:apedComparison} also show that the $f/r$ ratio is a poor predictor of temperature for these low density, evolving plasmas. Even the theoretical curves for AtomDB 1.3 and 2.0 flatten and lose their predictive power above 2 keV.

\begin{figure}
\begin{center}
\includegraphics[angle=0,width=\columnwidth]{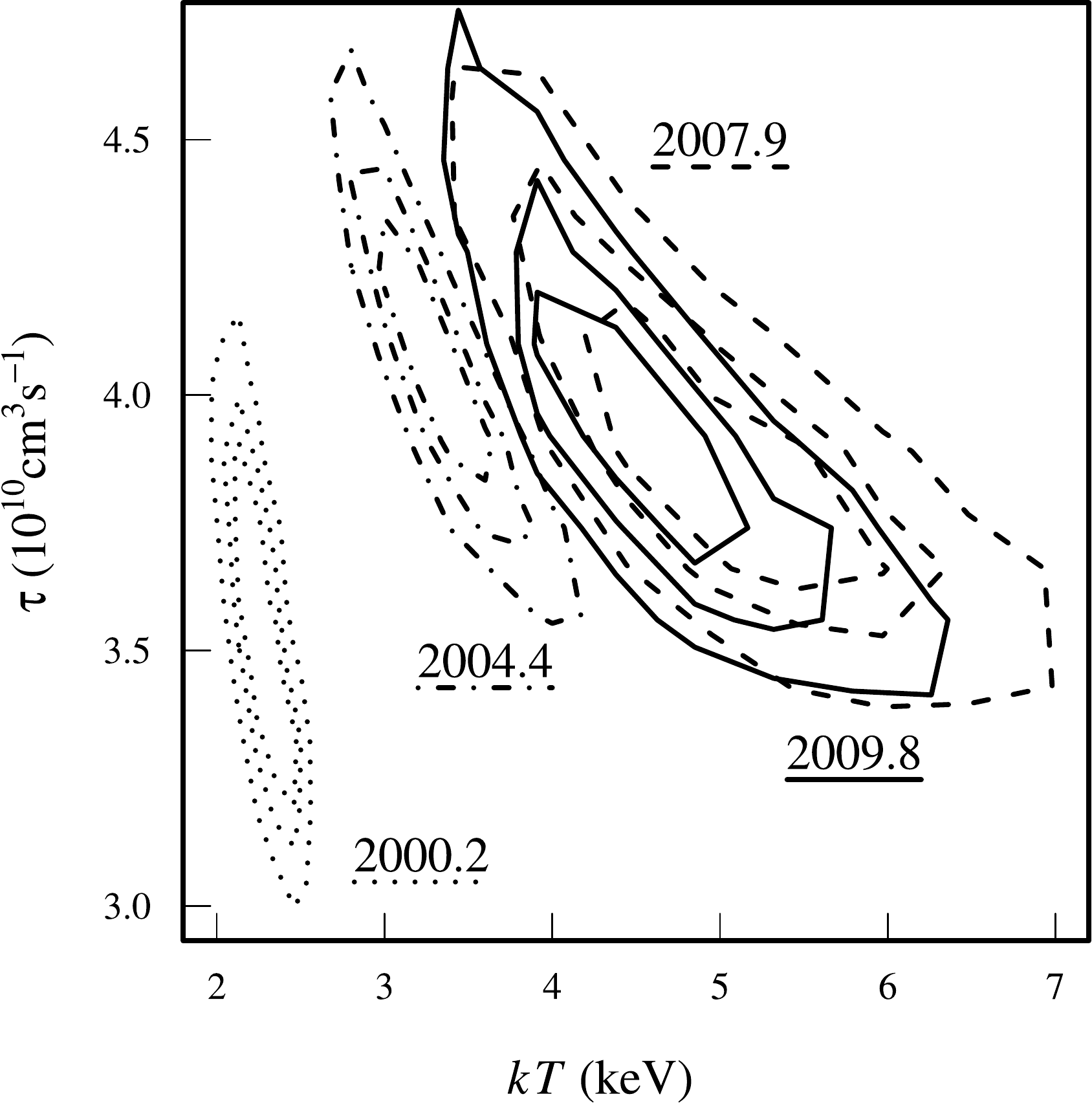} \\
\end{center}
\caption{
R02 shows distinct signs of evolving temperature.The confidence contours for the four epochs do not overlap in $kT$-$\tau$ space, indicating that the ACIS parameters have evolved. 
The three rings for each epoch indicate 68\%, 95\%, and 99\% confidence levels.
\label{fig:evolvingKnot}}
\end{figure}

\vspace{12pt}

\subsection{Plasma evolution}

On the whole, the selected knots show few signs of evolution, either in the derived plasma model parameters or the dispersed Si line ratios. 

The HETG results (Figure~\ref{fig:hetgvalues} and Table~\ref{tab:HETGresults}) yielded mixed trends. Six knots showed increasing H-like/He-like Si ratios (the plasmas ionized), while the ratio decreased significantly for only one knot (the plasma recombined). The $f/r$ ratios for most knots are consistent between the two epochs, within the errors.

The ACIS results (Figures~\ref{fig:acisresults1} -- \ref{fig:acisresults3} and Tables~\ref{tab:ACISresults} and \ref{tab:acistied}) revealed only five knots with evolving (i.e., non-overlapping) $kT$--$\tau$ contours. All evolving knots trended to higher temperatures over the decade of observation. Three knots in this evolving group showed evidence of aging (growing $\tau$), while one in fact cut its ionization age by a factor of four. As an example, Figure \ref{fig:evolvingKnot} exhibits distinctly non-overlapping confidence contours, which we identify as knot evolution.

\subsubsection{Evolution does not correlate with radius}

\begin{figure}
\begin{center}
\includegraphics[angle=0,width=0.75\columnwidth]{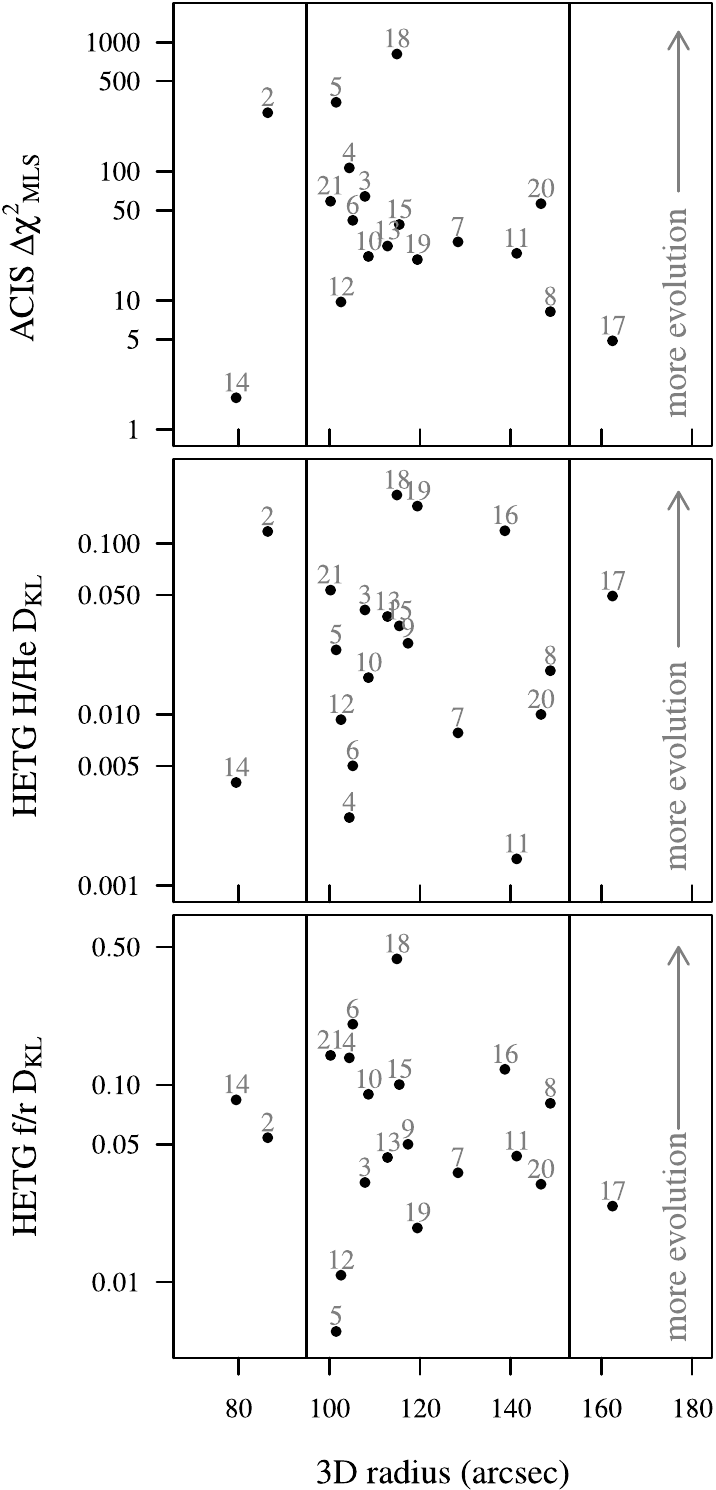} \\
\end{center}
\caption{The evolution of a knot does not correlate with its distance to the reverse shock. As a function of derived 3D radius (the $x$--axis), we plot three measures of evolution (the $y$--axis of the three separate plots). For the HETG parameters, we apply the K-L divergence, and for the ACIS values we employ a maximum likelihood swap. In both cases, higher values indicate more evolution. The reverse and forward shock (black vertical lines) are shown for comparison.
\label{fig:radiiEvol}}
\end{figure}

It is natural to ask if the few evolving knots are closer to the reverse shock than the others. To answer this question, we need first to locate the knots within the 3D structure of Cas A, then we must define a quantitative measure of evolution. 

In order to place the knots in 3D, the HETG measurements are essential: both the precise velocity measurements and the proper motions are needed to transform to physical coordinates with the appropriate scaling. We apply the prescription of \citet{DeLaney10} to effect the transformation from velocity space to position space, refining their technique to use individual knot expansion rates instead of the average. The results are presented in Table~\ref{tab:dopPropTable}.

\begin{deluxetable}{ccc}
\tablewidth{0pt}
\tablecaption{Doppler velocities and average radial\\ proper motions of the knots.
\label{tab:dopPropTable}}
\tablehead{ 	Region 	& Doppler velocity 	& Radial proper motion \\
					& (km s$^{-1}$)			& (milliarcseconds yr$^{-1}$)}
\startdata
R01 & -2340 & 38 \\  
R02 & -1360 & 221 \\  
R03 & 0 & 167 \\  
R04 & -150 & 134 \\  
R05 & -1450 & 210 \\  
R06 & -1110 & 192 \\  
R07 & 3780 & 125 \\  
R08 & 2740 & 140 \\  
R09 & -540 & 146 \\  
R10 & 1670 & 213 \\  
R11 & 750 & 66 \\  
R12 & -90 & 265 \\  
R13 & -810 & 249 \\  
R14 & 4300 & 279 \\  
R15 & 1070 & 279 \\  
R16 & -1360 & 313 \\  
R17 & -1710 & 307 \\  
R18 & -850 & 112 \\  
R19 & -880 & 166 \\  
R20 & -2300 & 109 \\  
R21 & 680 & 229
\enddata
\end{deluxetable}

Evolution is quantified differently for the HETG and ACIS results, due to the different statistics employed. Since the error bars in the HETG analysis are interpreted as widths of probability density functions, the Kullback-Leibler divergence ($D_{KL}$) can be used \citep{Kullback:1951cg}. The KL divergence measures the level of similarity between two distributions in an information theoretic manner. In Figure \ref{fig:radiiEvol}, higher $D_{KL}$ values indicate that the probability distributions of the 2010.3 parameters differ more from those in 2001.4. The ACIS confidence contours are not probability distributions, so we use a different measure of evolution: what we call a ``maximum likelihood swap'' (MLS). We substitute the best fit values of the 2009.8 epoch for the 2000.2 parameters, then evaluate the $\chi^2$. The resulting difference, $\Delta\chi^2_{MLS}$, corresponds to the probability of the true 2000.2 parameters being as extreme as the 2009.8 ones.

Our study does not reveal any correlation between evolution and distance from the reverse shock, which is shown in Figure \ref{fig:radiiEvol}. No correlations with norm or $\tau$ were detected, either (not shown). A complicated shock morphology may underlie this observed lack of correspondence. That is, any smaller scale secondary shocks could prematurely age some areas. The 3D morphology findings of \citet{DeLaney10} reveal many asymmetries, for instance, which may affect shock propagation. 

\subsection{ACIS and HETG cross-check}

\begin{figure}
\begin{center}
\includegraphics[angle=0,width=\columnwidth]{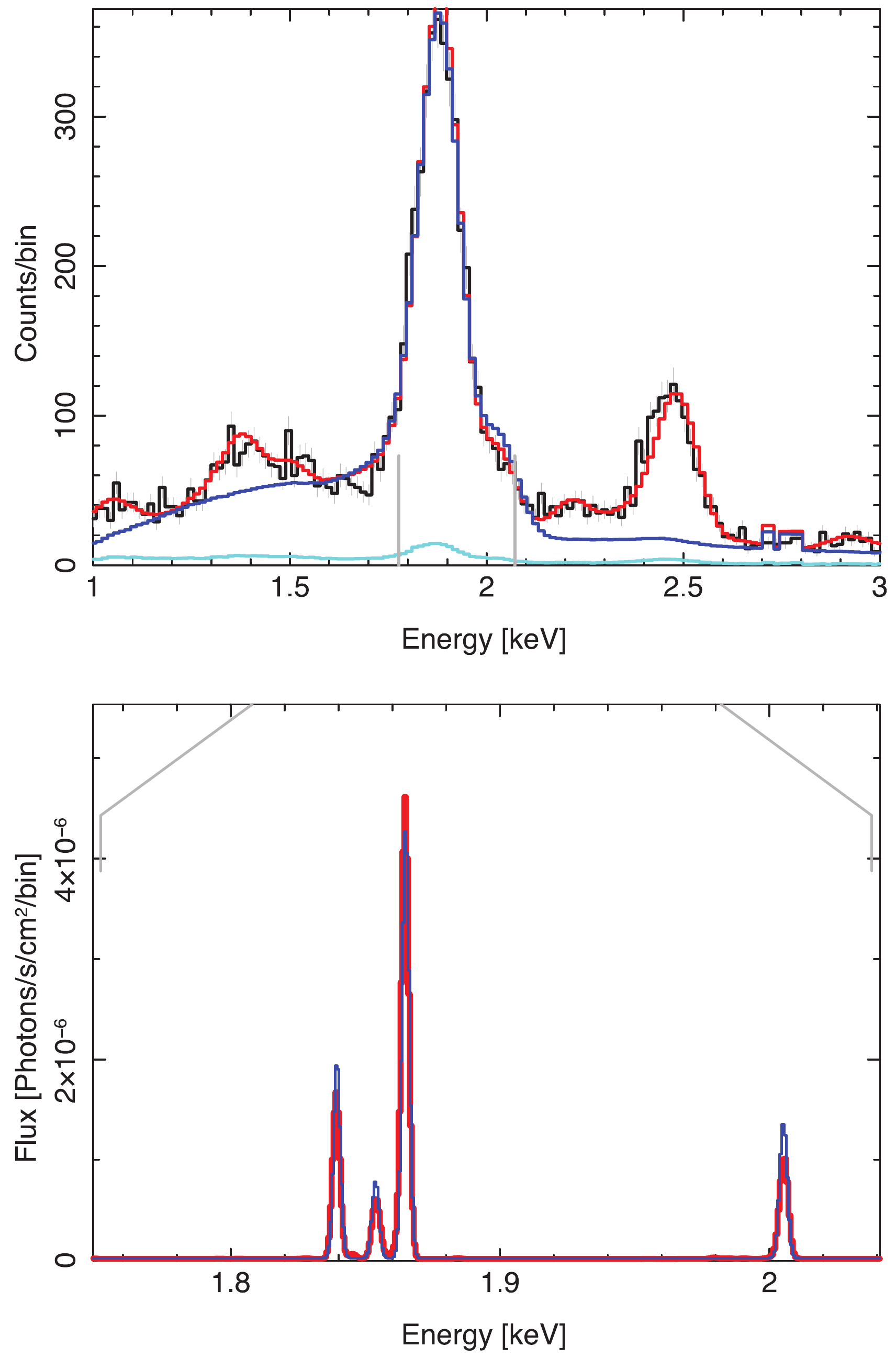} 
\end{center}
\caption{The measured HETG H/He-like ratios for the Si line agree well with the prediction from the best fit \vnei\ model from the ACIS analysis. Top: the ACIS data (black), \vnei\ model (red), and HETG 4-gaussian model, scaled, folded through the ACIS response, and with a bremsstrahlung component added (blue) agree in the Si region for R13. (ACIS background model is in light blue.) Bottom: a zoom-in of the underlying Si lines shows remarkable consistency between the HETG data (black), HETG model (blue), and scaled \vnei\ model (red). The lines are, left to right: Si XIII f, Si XIII i, Si XIII r, and Si XIV. 
\label{fig:H-He-comp-R13}}
\end{figure}

\begin{figure}
\begin{center}
\includegraphics[angle=0,width=\columnwidth]{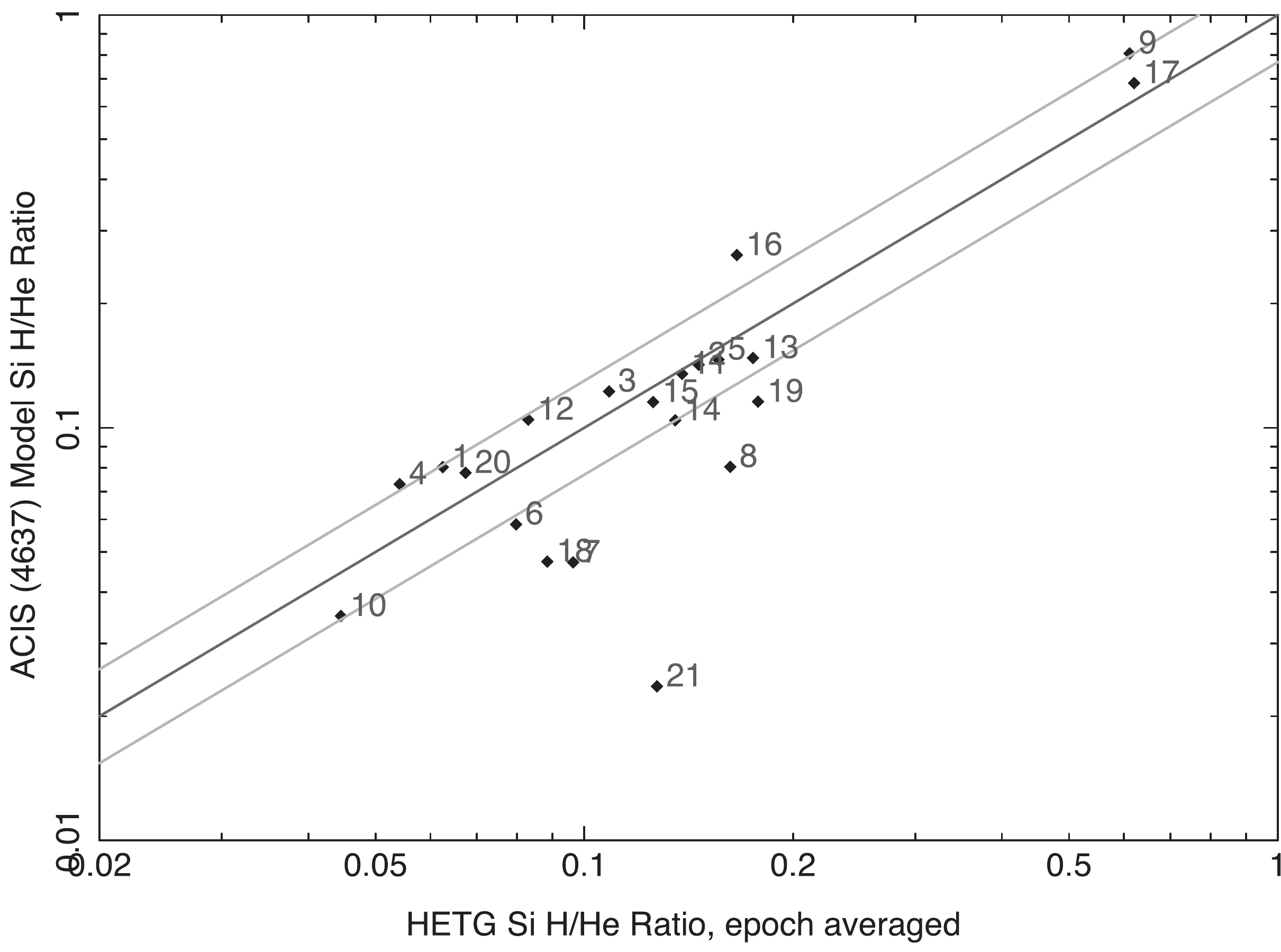} 
\end{center}
\caption{The quantified comparison of Figure~\ref{fig:H-He-comp-R13} shows good agreement between the HETG and ACIS models for most knots. The 30\% assumed systematic errors in the HETG measurement are shown with the top and bottom solid lines.
\label{fig:H-He-comp}}
\end{figure}

The results of the two different analyses are consistent with each other. The ACIS spectral response cannot resolve the He-like triplet lines, but it can separate the Si XIII triplet (as a single blend) from the Si XIV Ly-$\alpha$. As a cross-check of the ACIS and HETG results, we looked at the measured Si XIV / Si XIII ratio from HETG, and the predicted ratio from the ACIS \vnei\ model (Figure~\ref{fig:H-He-comp-R13}). The results, in Figure~\ref{fig:H-He-comp}, show consistency for most knots within 30\%. The effective area uncertainties for the \chandra\ instruments are on order of 10\%.

\section{Discussion}
\label{sec:disc}

\subsection{The lack of measured evolution\\ in the knots' plasma states}

The new HETG observation was intended to catch the knots during evolution, motivated by the high electron density of the knots inferred from \Laz.  Combining densities on the order of $100\mbox{ cm}^{-3}$ with a baseline of $\sim$\,9 years predicts a change of $\Delta\tau \approx 2.8\times 10^{10}\ccs$, which would produce observable changes in the plasma through its line ratios and $\tau $ values. As demonstrated above, neither the \vnei-modeled temperatures and ionization ages nor the HETG line ratios show changes of this magnitude. (Cf. Figure~\ref{fig:evolvingKnot}, which shows a spread of only\ $ 1.5 \times 10^{10}\ccs$. For even the low resolution ACIS spectral model, forcing $\tau$ to a higher value results in obvious disagreement with the data, due to an increased Si XIV/XIII ratio.) We must, therefore, look into the various assumptions in our modeling to explain this lack of evolution.

\subsubsection{Explaining the non-evolution with chemical makeup} 
\label{sec:vneidegen}

\begin{deluxetable*}{ccccccc}[h]
\tablewidth{0pt}
\tablecaption{Derived electron densities under various assumptions of the knot components. (Section~\ref{sec:vneidegen})
\label{tab:neTable}}
\tablehead{Region & Volume & $\tau$ (2004.4) & $n_e$ (90\% limits) &$n_e$ (nominal) &  $n_e$ (CN-rich) & $n_e$ (Si-rich)\\
		& ($10^{50} \mbox{ cm}^3$) & ($10^{10}\mbox{ s cm}^{-3}$) & ($\mbox{cm}^{-3}$)& ($\mbox{cm}^{-3}$)& ($\mbox{cm}^{-3}$)& ($\mbox{cm}^{-3}$)}
\startdata
R01 & 21.7 & 3.81  & 35.0 & 286.6 & 118.1 & 15.5 \\  
R02 & 12.4 & 4.05  & 51.2 & 280.9 & 116.0 & 17.2 \\  
R03 & 15.1 & 4.97  & 21.8 & 338.0 & 139.1 & 16.1 \\  
R04 & 102.3 & 4.60 & 50.8 & 239.8 & 98.7 & 11.7 \\  
R05 & 5.8 & 4.53   & 57.1 & 413.1 & 169.9 & 19.0 \\  
R06 & 7.7 & 3.49   & 65.7 & 263.1 & 108.5 & 15.1 \\  
R07 & 24.5 & 2.93  & 30.7 & 155.8 & 64.8 & 12.8 \\  
R08 & 7.5 & 4.03   & 45.2 & 248.4 & 102.8 & 17.1 \\  
R09 & 7.2 & 20.56  & 1105. & 360.1 & 149.3 & 26.7 \\  
R10 & 9.6 & 3.58   & 181. & 191.5 & 79.3 & 13.3 \\  
R11 & 17.7 & 5.79  & 46.2 & 274.3 & 113.0 & 14.5 \\  
R12 & 14.1 & 4.02  & 54.8 & 260.8 & 107.4 & 13.7 \\  
R13 & 5.2 & 7.39   & 609. & 323.6 & 133.6 & 19.7 \\  
R14 & 9.1 & 4.73   & 57.8 & 313.1 & 129.0 & 16.7 \\  
R15 & 7.0 & 3.73   & 211. & 244.0 & 100.7 & 14.8 \\  
R16 & 12.7 & 12.55 & 173. & 267.9 & 110.4 & 14.2 \\  
R17 & 16.2 & 10.37 & 333. & 107.1 & 45.0 & 11.2 \\  
R18 & 4.3 & 10.20  & $-$499* & 615.4 & 253.5 & 32.4 \\  
R19 & 6.1 & 8.59   & 903. & 327.3 & 135.1 & 19.7 \\  
R20 & 60.1 & 3.31  & 28.7 & 193.8 & 80.0 & 11.4 \\  
R21 & 8.1 & 2.40   & 38.6 & 105.5 & 43.6 & 7.2
\enddata
\tablenotetext{*}{R18 underwent ``recombination'', with its $\tau$ values decreasing over the decade. Therefore, the simple aging model can provide no information about the density.}
\end{deluxetable*}

As we will show, under our original model assumptions, the most optimistically low estimates for the knots' electron densities cannot explain their lack of spectral evolution. However, when we relax these assumptions about the elemental makeup of the knot, we find a physical state that could explain the observed spectral stasis.

We can turn our limits on $\Delta \tau$ into an upper limit on the electron density, since the plasma evolution depends on $n_e$, for a given time baseline.  A very generous upper limit is given by considering a $\tau $ change given by the difference between the epoch 2009.8 90\%\ upper-limit $\tau$ value and the epoch 2000.2 90\%\ lower-limit value.  Dividing this $\Delta\tau$ by the time baseline gives the ``$n_e$ (90\% limits)'' values in Table~\ref{tab:neTable}, which are in the 10's to 100's of cm$^{-3}$.  

The conversion of \vnei\ parameters to plasma density is straightforward, although not typically implemented in the modeling software.  \Laz\ outlines the technique (and we detail the calculation in Appendix~\ref{sec:ACIS-anal}): approximate the average number of electrons stripped per ion, then convert that to an absolute density through factors of the \vnei\ abundances and emission norm, by approximating the knot as a uniform density spheroid. The 2004.4 epoch model parameters are used to produce the densities ``$n_e$ (nominal)'' given in Table~\ref{tab:neTable}.  The values are too high above the upper limit densities.

The straightforward calculations of $n_e$ yielded values too high, so we must find out how to reduce the density while maintaining spectral consistency. Our derivation of the electron density leaves us with only a few slightly adjustable parameters to tune: 

\[ n_e \propto \sqrt{\frac{X_H \times n_e /n_H }{f} } ,\] namely $X_H$ -- the \vnei\ abundance of hydrogen relative to solar, $n_e /n_H$ -- the number of electrons per hydrogen ion in the plasma, and $f$ -- the filling fraction of the material in the emitting volume. (The uncertainties of the distance measurements to Cas A translate to only a 4\% error in the density, so we neglect that parameter.) We can start by changing the first two parameters with a different chemical makeup for the knot.

The low-Z elements provide the continuum, but the model cannot distinguish between their contributions. During fitting, therefore, we froze C, N, and O to 5 times solar (Appendix~\ref{sec:ACIS-anal}), and H and He to 1. We can investigate alternate abundance sets of low-Z elements that produce the same continuum, yet lead to different values of $n_e$. For instance, higher Z elements have more electrons to give up, so can compensate for a lower abundance of H. 

As an example of this continuum degeneracy, we consider a predominantly C and N makeup. (Cas~A exhibits little emission from C or N, but we care about the lack of H and He in this straw man model.) Changing the \{H, He, C, N, O\} abundances from \{1, 1, 5, 5, 5\} to a C- and N-rich plasma with \{0.002, 0.20, 45, 45, 5\} will produce a spectral model with the same norm, but a reduced $n_e$. (We increase C and N, but not O, based on the results of \citet{Dewey:2007ue}, though a N-rich plasma could do the same job.) Tweaking the abundances this way increases $n_e /n_H$ by about 100, partially undercutting the benefit of a lower $X_H$. The resulting $n_e$ values (``$n_e$ (CN-rich)'' column in Table~\ref{tab:neTable}) are reduced by over a factor of 2, but remain comfortably above the upper limits. This chemical makeup cannot explain the lack of evolution. 

We discard the CN-rich model, but the above example of an enriched metal plasma motivates an extreme scenario:  a plasma constituted only of elements with $Z>8$. Such a composition places a lower limit on the electron density that can still reproduce the observed line fluxes. Densities calculated with each knot's individual best-fit abundances -- but with $Z\leq 8 = 0$ -- fall below the estimated ``90\%\ limits'' densities (Table~\ref {tab:neTable}, ``Si-rich''), in the range of 10--20 cm$^{-3}$. The lack of evolution for many knots could therefore be explained by a low density, Si-rich plasma component.

To make this metal-rich knot model more physical -- and spectrally consistent -- we need only split our nominal knot into two components. For both components, we use the same parameters as the nominal knot: the 2004.4 maximum likelihood $kT$, $\tau$, and abundances (with the fiducial continuum set \{H, He, C, N, O\} $=$ \{1, 1, 5, 5, 5\}). The metal-rich component is given an 85\% fill fraction, and the elements with $Z\leq8$ are assigned contamination-level abundances of $10^{-2}$ solar. Complementarily, the low-Z component fills 15\% of the knot to bolster the continuum, and has $Z>8$ abundances set to contamination level.  

The two models with complementary abundance sets recover the same well fitting spectral model as the nominal case (Figure~\ref{fig:knot-cartoon}). Unlike the CN-rich case, the altered chemical composition results in slightly higher $n_e /n_H$, while $X_H$ is cut by a factor of 100. The resulting $n_e$'s for the metal-rich components -- which contain the Si we have been using as a tracker of evolution -- are below the upper limit values (Table~\ref {tab:neTable}), consistent with the non-evolution we observe. 

\subsubsection{Evidence for metal dissociation}

A model that dissociates high-Z metals from the other elements can produce low $n_e$, thereby explaining the lack of evolution. Though we found no other simple solution to regain self-consistency, we acknowledge the model presented here may not be the only answer. 

\begin{figure}
\begin{center}
\includegraphics[angle=0,width=\columnwidth]{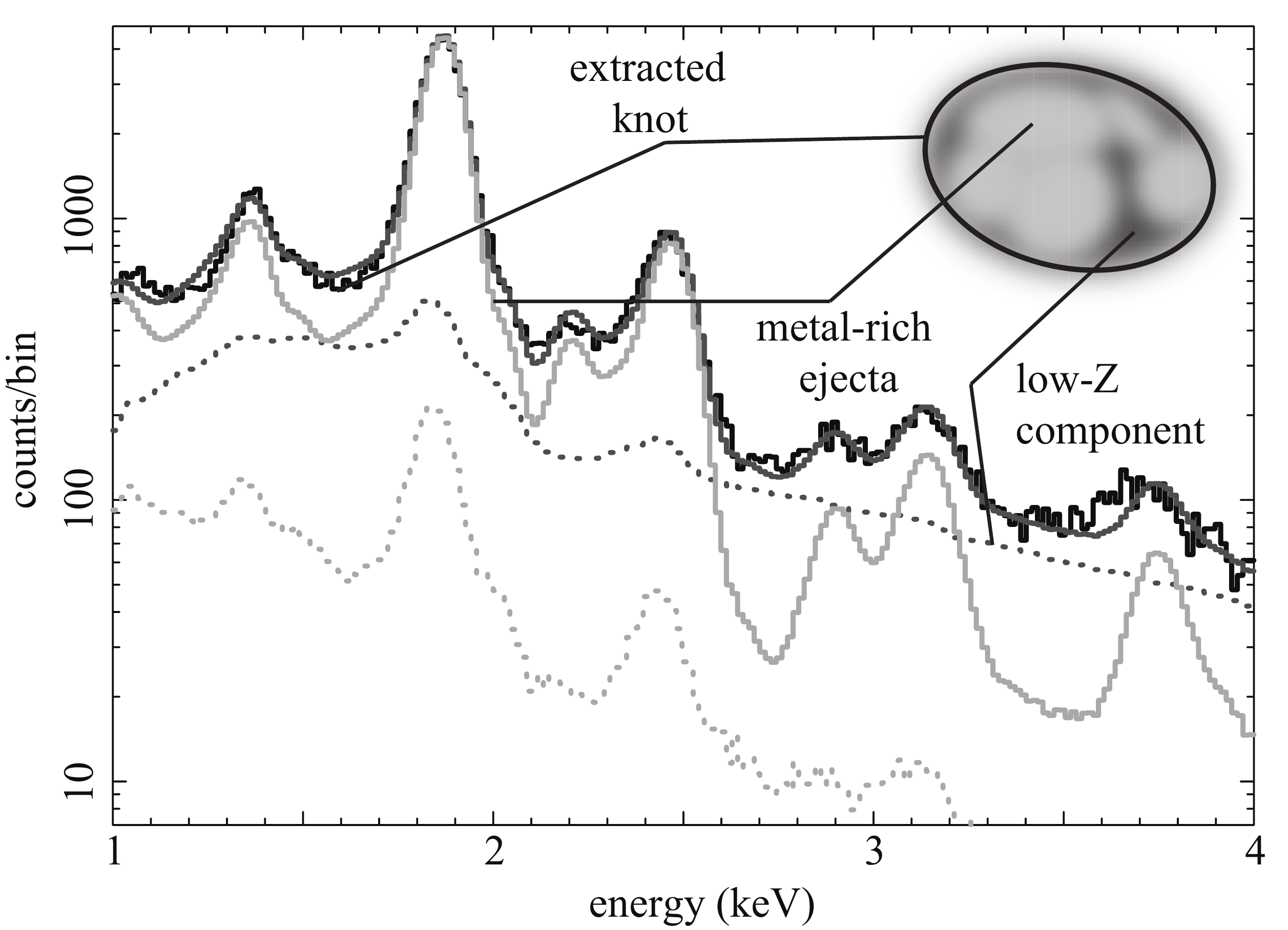} 
\end{center}
\caption{The simple model of a knot split into two spectral components -- a high metallicity finger expanding into a denser low metallicity plasma -- reproduces the observed spectrum. The observed knot spectrum (black) is a sum (dark grey) of the low metallicity component (dotted dark grey) and the high metallicity component (light grey). The background spectrum (dotted light grey) does not resemble the low metallicity component. Metal dissociation within a knot could explain why so little evolution was observed over the decade. 
\label{fig:knot-cartoon}}
\end{figure}

This dissociated plasma is made up of low density pure ejecta and a low-Z component. The low-Z ($Z \leq 8$) plasma provides the continuum emission we see and has higher density: the $n_e$'s range from 300--1600 cm$^{-3}$, which correspond to radiative cooling times greater than 1900 years. This low-Z plasma would be physically separated from a less dense, ejecta-rich ($Z > 8$) plasma that accounts for the observed line emission. These plasmas share the same $kT$ and $\tau$, by construction, so the dual model sums to single-component \vnei\ model, reproducing the good spectral fits (Figure~\ref{fig:knot-cartoon}). Only the interpretation of the physical picture is changed: the components are unmixed. 

The data in this work cannot constrain the specific physical model for knot dissociation, but hydrodynamical simulations offer indications. We envision a low density parcel of pure high-Z ejecta pushing into a denser layer of low-Z material ahead of it, much like the tip of a Rayleigh-Taylor (RT) finger. The high-Z component, though, could manifest as an expanding clump implanted into the low-Z plasma, or be turbulently mixed on finer scales.

The scales for this dissociation would be small, if the components are not separated along the line of sight. \cite{Lopez:2011kb} found evidence for well mixed ejecta in Cas A, down to $3^{\prime\prime}$ ($\sim 10^{17}$ cm). Likewise, our knot-defining algorithm did not see large spectral differences at the pixel scale. Investigating infrared knots, \citet{isensee12} found a larger scale (0.1 pc, $5^{\prime\prime}$) line-of-sight separation between the O and Si layers in one $40^{\prime\prime}$ by $40^{\prime\prime}$ region of Cas A. These knots are more recently shocked than the X-ray knots, and do not admit direct comparison. Our proposed model would require dissociation as a line-of-sight effect, or at scales unresolved by the \chandra\ optics, which have a PSF matched to the pixel size, $0.5^{\prime\prime}$ ($\sim 2.5 \times 10^{16}$ cm). 

Early-times explosion models consistently show large scale RT mixing between burning layers, though just barely resolve the small scales of our knots. \citet{Hammer10} found strong mixing on large scales, with high-Z bullets probing deep into the outer hydrogen layer. The smallest structures at 2.5 hours span $10^{11}$ cm, roughly the size of these X-ray knots at that time. The work of \citet{Joggerst10} found that mixing can persist longer ($\sim 1$ day after the explosion). The rapid expansion of the remnant may leave elemental mixing incomplete, though to date there have been no studies to determine on what scale (private communication, H. Janka, 2011).  

In the late-stage mixing simulations, several enhancements to the RT mixing have been suggested to force the fingers through the whole intershock region, where our Si-rich knots lie. RT instabilities could reach to the forward shock with the help of high enough compression ratios \citep{blondin01b}, large, underdense Fe bubbles colliding with the reverse shock \citep{blondin01a} or high density contrasts in the originating ejecta clumps \citep{orlando12}. \citeauthor{orlando12} observe density contrasts of 3 to 4 orders of magnitude over 0.5 pc (for a 1000 yr old simulated remnant), while we claim contrasts of just over 1 order of magnitude, spanning $10^{-3}$ pc in Cas A. 

The recent work of \citet{ellinger12} aims to bridge the early and late epochs. Their simulations have evolved Cas A to 30 years, but have revealed little mixing between the Si and O layers, in contrast with \citet{Hammer10} and \citet{Joggerst10}, who found more Si in the RT fingers. We await their extended simulations to compare with our dissociated metal hypothesis. 

It is worth noting that our dissociated model remains simple, with a shared temperature and ionization age. If the ejecta are not fully mixed at these small scales, we would expect the two components to have different temperatures and ionization ages, though we achieve good spectral agreement without turning these additional knobs. This additional freedom could reconcile our differences with \Laz's $kT$ and $\tau$ values. (Initial investigations show that, if freed, $kT$ and $\tau$ for the high-Z component would not change much, while the dissociated model permits a much wider range of low-Z $\tau$ values.)

If this dissociation is the answer to the non-evolution riddle, it could inform elemental mixing during supernova explosions. 

\subsubsection{Flux evolution}

The brightened knot, R21, which we added to the set from \Laz, was also investigated in \cite{Patnaude07}. They too claimed spectral evolution in R21, with $kT$ increasing from 1.1 to 1.5 keV from 2002 to 2004, and $\tau$ increasing from $4.6 \times 10^{10}$ to $53.0 \times 10^{10}\mbox{\ s\ cm}^{-3}$. From 2000 to 2004, our results show a similar increase in temperature, but $\tau$ is not only lower ($\sim 2 \times 10^{10}\mbox{\ s\ cm}^{-3}$), but fairly constant over that period. The disparities may arise from different model assumptions: both analyses use an NEI model with variable abundances, but \cite{Patnaude07} fit the spectrum over a smaller energy range, did not tie parameters across epochs, and derived only single-parameter confidence intervals.

In general, for many of the knots, the brightness of the Si lines changed dramatically -- up to a factor of 3 --  over the decade, as Figure \ref{fig:hetgvalues} shows. However, a large-area, broadband investigation of Cas~A by \cite{Patnaude11} showed only the slightest decrease in 1.5 -- 3.0 keV flux in the same period. Evidently, the uncorrelated variations in the individual knots average out over time and position, giving the impression of steady thermal emission that \citeauthor{Patnaude11} observe. 

\bigskip
\bigskip

\subsection{Model degeneracies}

We have found that the \vnei\ parameters \emph{kT} and $\tau$ are often anti-correlated. A higher \emph{kT} and lower $\tau$ can produce a spectrum with equal statistical likelihood to one with lower \emph{kT} and higher $\tau$. The characteristic of this degeneracy is a banana-shaped confidence contour, such as in Figure  \ref{fig:evolvingKnot}. 

There is a physical justification for the strong anti-correlation between $kT$ and $\tau$. The equilibrium ratio of the Si H-like Ly-$\alpha$ to the He-like triplet for a plasma is monotonically increasing with temperature. As a NEI plasma evolves toward CIE, this ratio also monotonically increases with $\tau $. Thus, for a given spectrum, models with higher $kT$ and lower $\tau $ or models with lower $kT$ and higher $\tau$ can result in the same likelihood.

This degeneracy explains why the discrepancies in the $kT$ and $\tau$ values derived in \Laz\ and in this analysis are anti-correlated.

Had we not used 2-dimensional confidence contours for the 2 \vnei\ parameters of interest, we would have come to different conclusions about the data, due to problems that arise with the use of 1-dimensional confidence intervals for multidimensional data. First, during the interval generation for one parameter, the other parameters of interest are treated as nuisance parameters, which is wrong from a mathematical perspective (the incorrect $\chi^2$ distribution is used). Second, the use of a $\chi^2$ distribution with fewer degrees of freedom underestimates the parameter uncertainty. Moreover, any underlying correlations are lost, leaving the analyst susceptible to false trends in best fit values, when in fact those points may live in the same ``banana.''

\section{Conclusions}

We have performed a detailed analysis of two HETG and four ACIS observations of Cas A spanning a 10-year baseline. A new set of detailed plasma kinematic and temperatures measurements has been presented. Good agreement was found for the HETG-- and ACIS--derived temperatures for about half of the knots. The outliers are well described by a plasma far from collisional ionization equilibrium, the state which is assumed for the HETG line-ratio temperature estimates. 

The low $\tau$ values and the high electron densities derived from this analysis predict a significant amount of plasma evolution in the 10-year baseline, in stark contrast with the observations. We propose a physical model of two plasmas -- one high metallicity, one low metallicity -- that remain unmixed at small spatial scales. The Si emission comes from the pure heavy metal ejecta (Ne through Fe), while the continuum is provided by the low density component, rich in the elements C, N, and O. This model fits the data well and explains the lack of evolution, since it requires a much lower electron density for the Si plasma and, therefore, a longer timescale for evolution. If validated, this model could place strong constraints on turbulence in supernova explosion models.

\vspace{6pt}
\acknowledgments

Support for this work was provided by NASA through the Smithsonian Astrophysical Observatory (SAO) contract SV3-73016 to MIT for Support of the Chandra X-Ray Center (CXC) and Science Instruments. CXC is operated by SAO for and on behalf of NASA under contract NAS8-03060.  Support was also provided by NASA award NNX10AE25G, NSF PAARE grant AST-0849736, the NASA Earth and Space Science Fellowship, the NASA Harriet Jenkins Pre-doctoral Fellowship Program, the Vanderbilt Provost Graduate fellowship, and the Japan Society for the Promotion of Science Fellowship.


\clearpage

\appendix
\section{Appendix A: Results Tables and Figures}
\label{sec:result-app}
The tables and figures in the following pages present the results of our kinematic and thermodynamic analysis of 21 knots in Cas A. 

\begin{figure}
\begin{center}
\includegraphics[angle=0,width=6in]{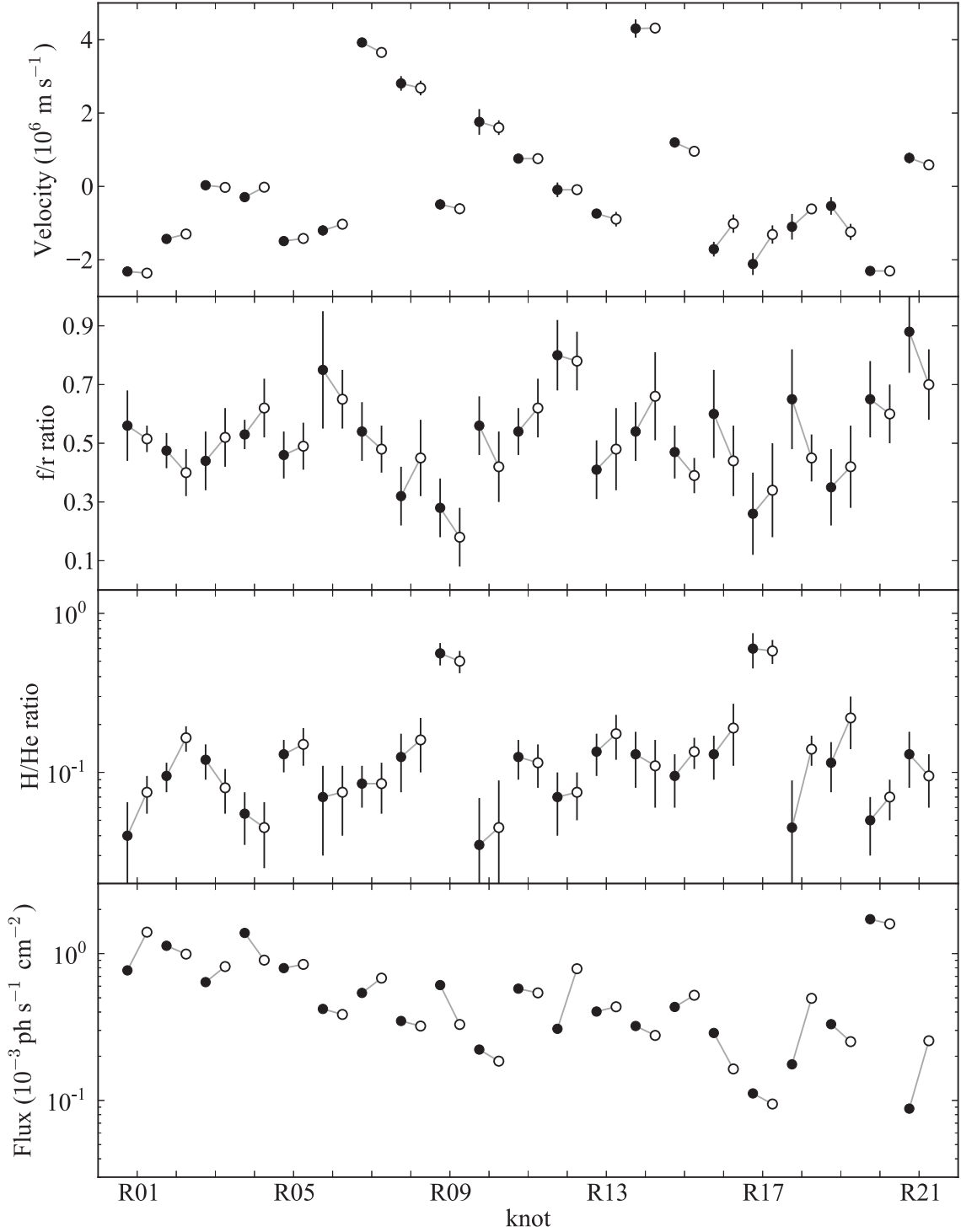} \\
\end{center}
\caption{The HETG measurements did not show much change over the decade, besides the norm values. The left-most point for each knot (black) is the result for the first epoch (2001) and the right-most point (white) the result for the second epoch (2010). The ``f/r'' label refers to the forbidden/recombination line ratio for Si XIII. The ``H/He'' label refers to the (Si Ly-$\alpha$)/(Si He-like triplet) line ratio. The norm errors were not computed.
\label{fig:hetgvalues}}
\end{figure}

Table~\ref{tab:HETGresults} lists the values extracted from the high-energy-resolution HETG analysis, and Figure~\ref{fig:hetgvalues} plots them. We extract the velocity of each knot, two line ratios, and the photon flux in the fitted line complex. The two line ratios are the ratio of the Si He-like triplet forbidden and recombination lines ($f/r$), which can only be obtained through the high-energy-resolution analysis, and the Si H-like to He-like ratio, which can be compared to the ACIS model fit value (shown in Figure~\ref{fig:H-He-comp}).

The inferred values of the plasma parameters of interest can be found in Table~\ref{tab:ACISresults}, and are plotted in Figures \ref{fig:acisresults1}, \ref{fig:acisresults2}, and \ref{fig:acisresults3} to show the correlations between $kT$ and $\tau$. 

\begin{deluxetable}{cccccc}[b]
\tablewidth{0pt}
\tablecaption{Measured Si line parameters\tablenotemark{a} for 21 Cas~A knots from 
spatial-spectral modeling of the HETG dispersed data.
\label{tab:HETGresults}}
\tablehead{ Region & Diam. & Velocity & Si XIII  & Si XIV         & Flux \\
         ~~--Epoch & ('')  & (\kms)   & (f/r)    &  (H/He ratio)  & (\,\tablenotemark{b}\,) }
\startdata
    R01--I & 3.00 & $ -2320 \pm 90 $ & $  0.56 \pm 0.12 $ & $  0.040 \pm 0.025 $  &  0.77 \\  
~~''~~--II & 4.80 & $ -2360 \pm 100 $ & $  0.52 \pm 0.05 $ & $  0.075 \pm 0.020 $  & 0.14 \\  
    R02--I & 4.80 & $ -1430 \pm 90 $ & $  0.48 \pm 0.06 $ & $  0.095 \pm 0.020 $  & 1.1 \\  
~~''~~--II & 5.04 & $ -1300 \pm 70 $ & $  0.40 \pm 0.08 $ & $  0.165 \pm 0.030 $  &  0.99 \\  
    R03--I & 3.72 & $ +30 \pm 100 $ & $  0.44 \pm 0.10 $ & $  0.120 \pm 0.030 $  &  0.64 \\  
~~''~~--II & 4.20 & $ -30 \pm 120 $ & $  0.52 \pm 0.10 $ & $  0.080 \pm 0.025 $  &  0.82 \\  
    R04--I & 5.76 & $ -290 \pm 100 $ & $  0.53 \pm 0.05 $ & $  0.055 \pm 0.020 $  & 1.4 \\  
~~''~~--II & 5.04 & $ -20 \pm 100 $ & $  0.62 \pm 0.10 $ & $  0.045 \pm 0.020 $  &  0.90 \\  
    R05--I & 3.48 & $ -1490 \pm 100 $ & $  0.46 \pm 0.08 $ & $  0.130 \pm 0.030 $  &  0.80 \\  
~~''~~--II & 3.36 & $ -1420 \pm 140 $ & $  0.49 \pm 0.08 $ & $  0.150 \pm 0.040 $  &  0.84 \\  
    R06--I & 3.60 & $ -1200 \pm 140 $ & $  0.75 \pm 0.20 $ & $  0.070 \pm 0.040 $  &  0.42 \\  
~~''~~--II & 3.60 & $ -1030 \pm 140 $ & $  0.65 \pm 0.10 $ & $  0.075 \pm 0.035 $  &  0.39 \\  
    R07--I & 6.00 & $ +3920 \pm 100 $ & $  0.54 \pm 0.10 $ & $  0.085 \pm 0.025 $  &  0.54 \\  
~~''~~--II & 5.04 & $ +3650 \pm 100 $ & $  0.48 \pm 0.08 $ & $  0.085 \pm 0.030 $  &  0.68 \\  
    R08--I & 3.36 & $ +2810 \pm 200 $ & $  0.32 \pm 0.10 $ & $  0.125 \pm 0.050 $  &  0.35 \\  
~~''~~--II & 3.60 & $ +2680 \pm 200 $ & $  0.45 \pm 0.13 $ & $  0.160 \pm 0.060 $  &  0.32 \\  
    R09--I & 3.84 & $ -490 \pm 140 $ & $  0.28 \pm 0.10 $ & $  0.560 \pm 0.090 $  &  0.61 \\  
~~''~~--II & 3.24 & $ -610 \pm 150 $ & $  0.18 \pm 0.10 $ & $  0.500 \pm 0.080 $  &  0.33 \\  
    R10--I & 4.80 & $ +1760 \pm 350 $ & $  0.56 \pm 0.10 $ & $  0.035 \pm 0.034 $  &  0.22 \\  
~~''~~--II & 4.80 & $ +1600 \pm 200 $ & $  0.42 \pm 0.12 $ & $  0.045 \pm 0.044 $  &  0.18 \\  
    R11--I & 4.92 & $ +760 \pm 120 $ & $  0.54 \pm 0.08 $ & $  0.125 \pm 0.035 $  &  0.58 \\  
~~''~~--II & 5.04 & $ +760 \pm 120 $ & $  0.62 \pm 0.10 $ & $  0.115 \pm 0.035 $  &  0.54 \\  
    R12--I & 3.48 & $ -90 \pm 200 $ & $  0.80 \pm 0.12 $ & $  0.070 \pm 0.030 $  &  0.31 \\  
~~''~~--II & 5.88 & $ -90 \pm 120 $ & $  0.78 \pm 0.10 $ & $  0.075 \pm 0.025 $  &  0.79 \\  
    R13--I & 3.60 & $ -740 \pm 120 $ & $  0.41 \pm 0.10 $ & $  0.135 \pm 0.040 $  &  0.40 \\  
~~''~~--II & 4.20 & $ -890 \pm 200 $ & $  0.48 \pm 0.14 $ & $  0.175 \pm 0.055 $  &  0.43 \\  
    R14--I & 4.32 & $ +4300 \pm 250 $ & $  0.54 \pm 0.10 $ & $  0.130 \pm 0.050 $  &  0.32 \\  
~~''~~--II & 4.08 & $ +4320 \pm 120 $ & $  0.66 \pm 0.15 $ & $  0.110 \pm 0.050 $  &  0.28 \\  
    R15--I & 3.72 & $ +1200 \pm 140 $ & $  0.47 \pm 0.09 $ & $  0.095 \pm 0.035 $  &  0.43 \\  
~~''~~--II & 4.20 & $ +960 \pm 120 $ & $  0.39 \pm 0.06 $ & $  0.135 \pm 0.030 $  &  0.52 \\  
    R16--I & 4.56 & $ -1710 \pm 200 $ & $  0.60 \pm 0.15 $ & $  0.130 \pm 0.040 $  &  0.29 \\  
~~''~~--II & 3.36 & $ -1020 \pm 250 $ & $  0.44 \pm 0.12 $ & $  0.190 \pm 0.080 $  &  0.16 \\  
    R17--I & 3.60 & $ -2110 \pm 300 $ & $  0.26 \pm 0.14 $ & $  0.600 \pm 0.150 $  &  0.11 \\  
~~''~~--II & 4.20 & $ -1310 \pm 250 $ & $  0.34 \pm 0.16 $ & $  0.580 \pm 0.100 $  &  0.095 \\  
    R18--I & 3.00 & $ -1100 \pm 350 $ & $  0.65 \pm 0.17 $ & $  0.045 \pm 0.044 $  &  0.18 \\  
~~''~~--II & 3.60 & $ -610 \pm 120 $ & $  0.45 \pm 0.08 $ & $  0.140 \pm 0.030 $  &  0.50 \\  
    R19--I & 3.60 & $ -530 \pm 240 $ & $  0.35 \pm 0.13 $ & $  0.115 \pm 0.040 $  &  0.33 \\  
~~''~~--II & 3.60 & $ -1240 \pm 220 $ & $  0.42 \pm 0.14 $ & $  0.220 \pm 0.080 $  &  0.25 \\  
    R20--I & 5.40 & $ -2300 \pm 80 $ & $  0.65 \pm 0.13 $ & $  0.050 \pm 0.020 $  & 1.7 \\  
~~''~~--II & 6.60 & $ -2310 \pm 100 $ & $  0.60 \pm 0.10 $ & $  0.070 \pm 0.020 $  & 1.6 \\  
    R21--I & 3.36 & $ +780 \pm 150 $ & $  0.88 \pm 0.14 $ & $  0.130 \pm 0.050 $  &  0.088 \\  
~~''~~--II & 3.36 & $ +590 \pm 100 $ & $  0.70 \pm 0.12 $ & $  0.095 \pm 0.035 $  &  0.26
\enddata
\tablenotetext{a}{These ratios and fluxes are ``un-absorbed'' based on the $N_H$
determined by ACIS fitting.}
\tablenotetext{b}{This un-absorbed flux is the sum of the four Si lines
in units of 10$^{-3}$~ph s$^{-1}$ cm$^{-2}$.}
\end{deluxetable}
\clearpage

\clearpage

\begin{figure}
\begin{center}
\includegraphics[width=.9\textwidth]{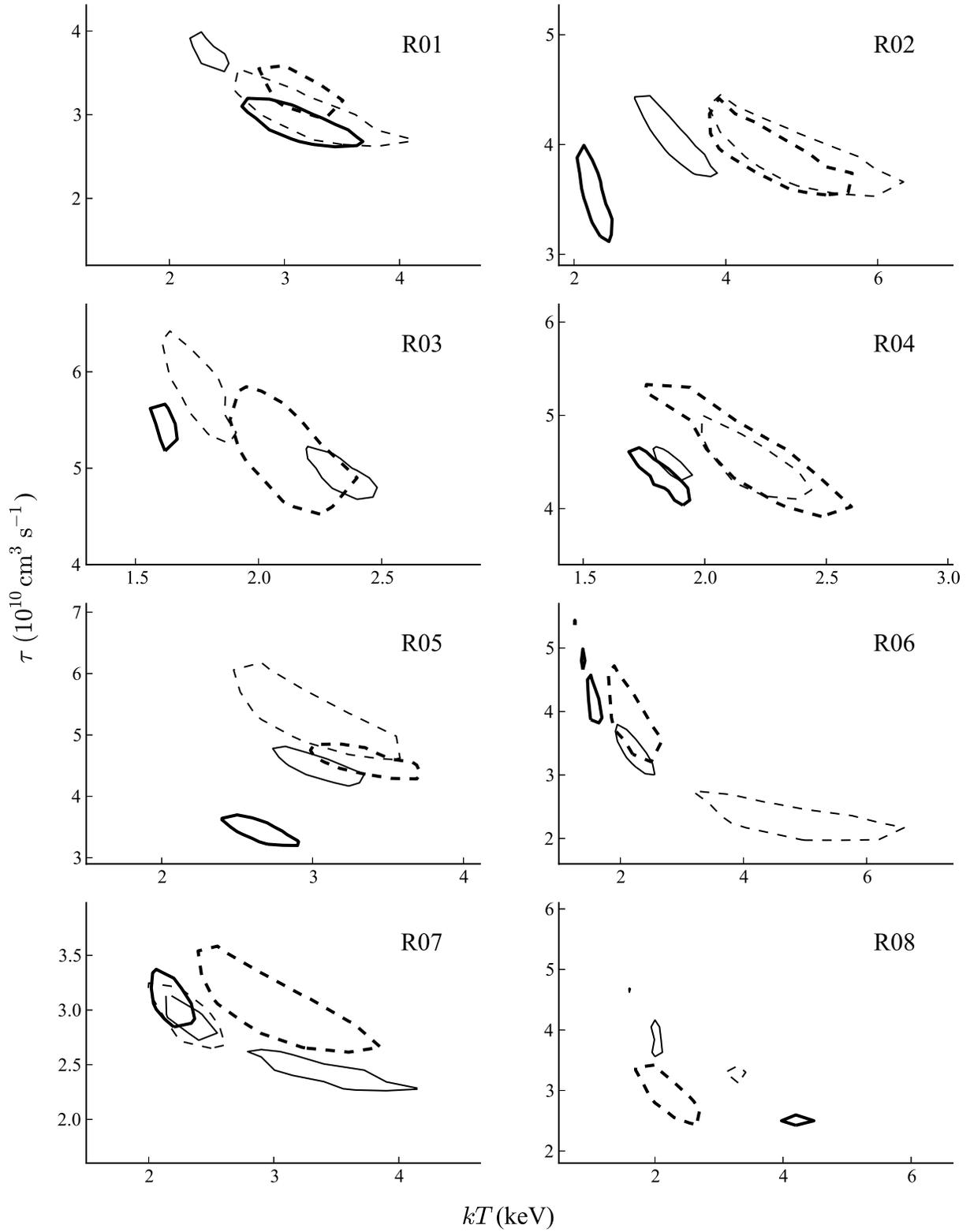} \\
\end{center}
\caption{The ACIS data likewise showed little evolution. The 95\% confidence contours in the $kT$--$\tau$ plane are shown for each epoch: 2000.2 (thick), 2004.4 (thin), 2007.9 (thin dashed), and 2009.8 (thick dashed). The joint fits between all four epochs are described in Appendix~\ref{sec:ACIS-anal}. Note the strong correlation between the two parameters.
\label{fig:acisresults1}}
\end{figure}
\clearpage
\begin{figure}
\begin{center}
\includegraphics[width=.9\textwidth]{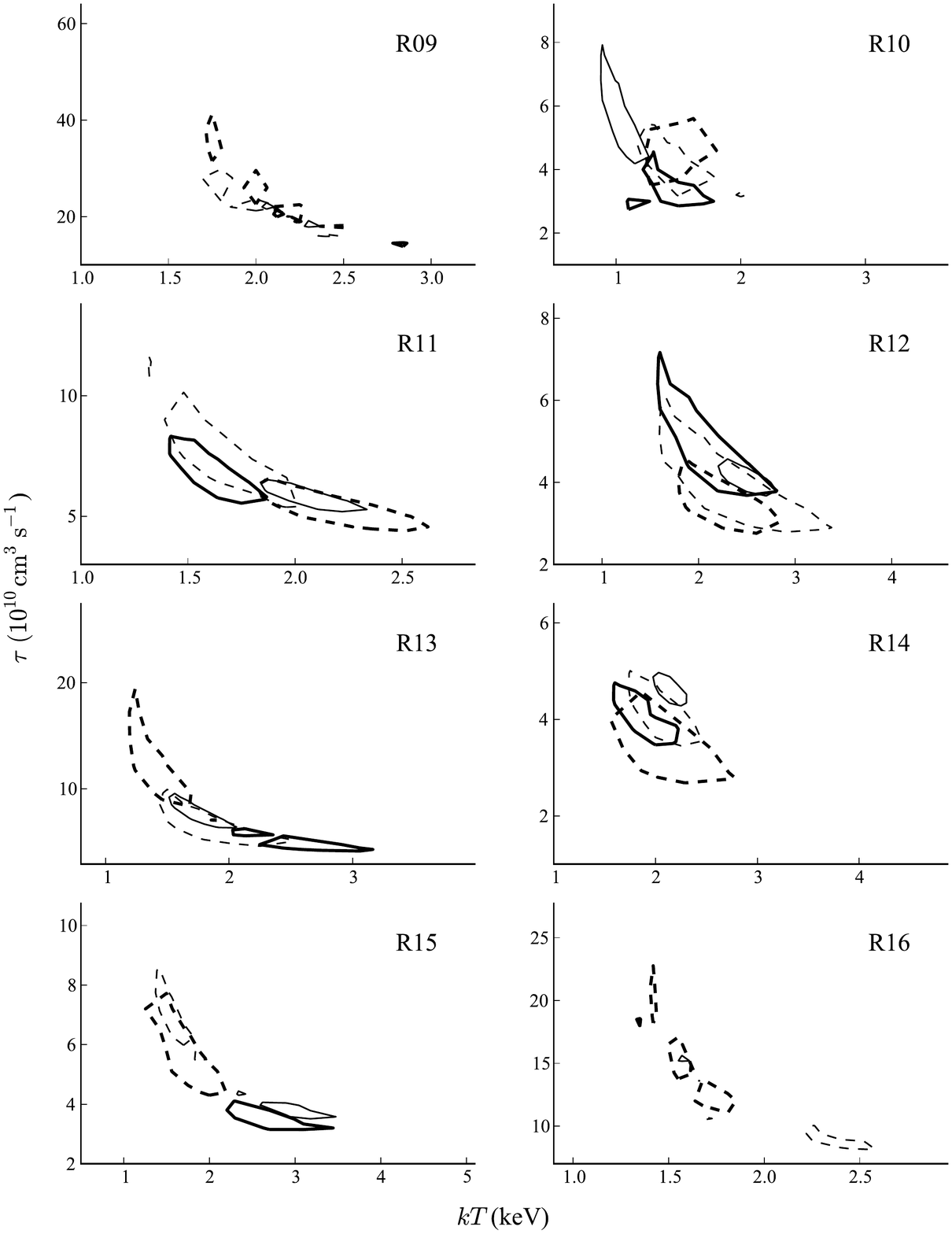} \\
\end{center}
\caption{Continuation of Figure~\ref{fig:acisresults1}. The 95\% confidence contours in the $kT$--$\tau$ plane are shown for each epoch: 2000.2 (thick), 2004.4 (thin), 2007.9 (thin dashed), and 2009.8 (thick dashed). 
\label{fig:acisresults2}}
\end{figure}
\clearpage
\begin{figure}
\begin{center}
\includegraphics[width=.9\textwidth]{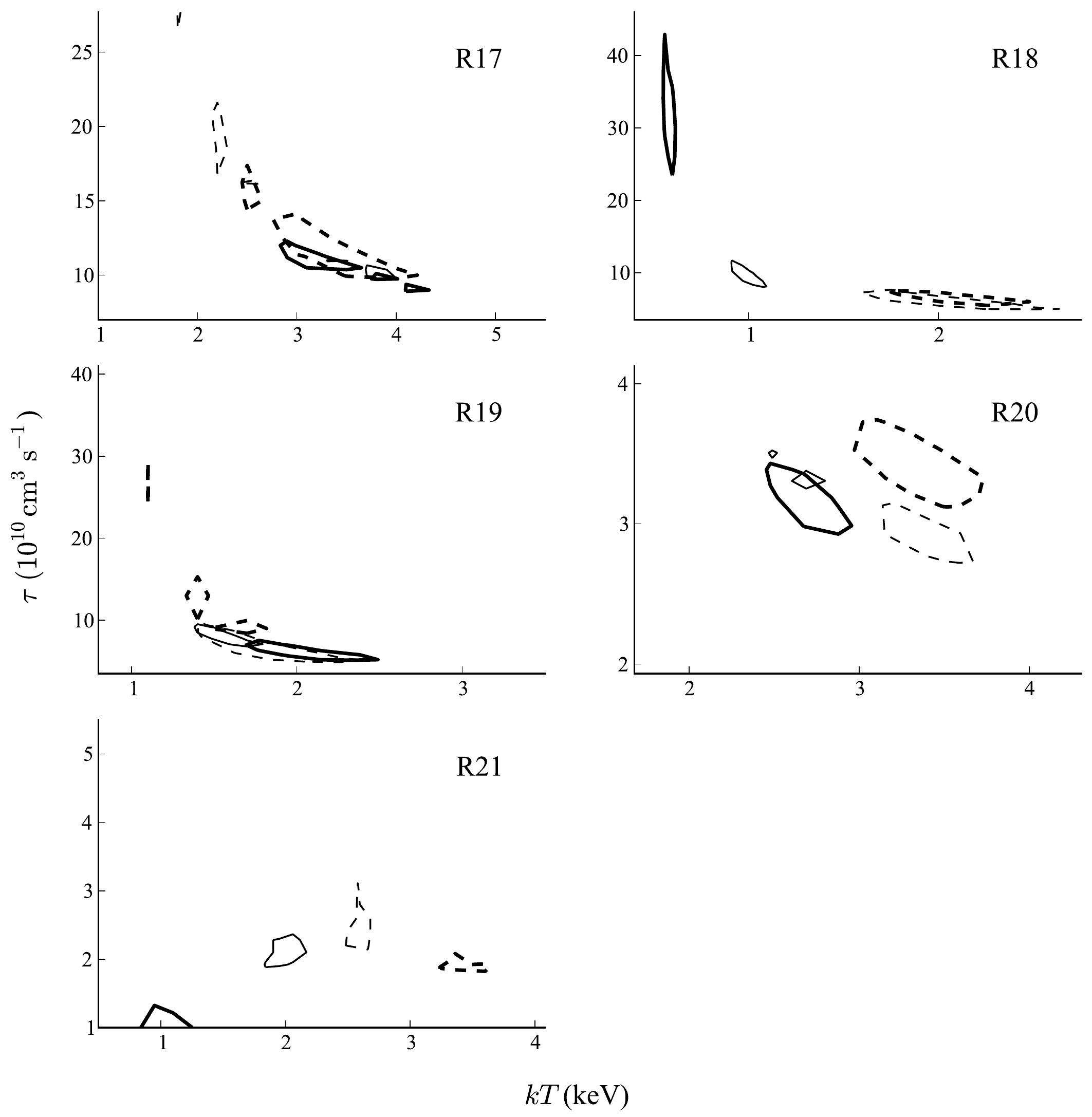} \\
\end{center}
\caption{Continuation of Figure~\ref{fig:acisresults1}. The 95\% confidence contours in the $kT$--$\tau$ plane are shown for each epoch: 2000.2 (thick), 2004.4 (thin), 2007.9 (thin dashed), and 2009.8 (thick dashed). 
\label{fig:acisresults3}}
\end{figure}
\clearpage

\begin{deluxetable}{cccccccccc}
\tablewidth{0pt}
\tablecaption{Limiting bounds of the 2-dimensional confidence $kT$--$\tau$ contours (90\%) from the ACIS analysis.
\label{tab:ACISresults}}
\tablehead{ Region & Epoch & $kT$ & $\tau$ & norm\tablenotemark{$\dagger$} &  Region & Epoch & $kT$ & $\tau$ & norm\tablenotemark{$\dagger$} \\
&   & (keV)   & ($10^{10} \mathrm{\ s\ cm}^{-3}$) & ($\times 10^{-4}$) &   &  & (keV)   & ($10^{10} \mathrm{\ s\ cm}^{-3}$) & ($\times 10^{-4}$)}
\startdata
\multirow{4}{*}{R01} & 2000.2 & 2.62 -- 3.68 & 2.6 -- 3.24 & 7.04  &\multirow{4}{*}{R12} & 2000.2 & 1.46 -- 2.87 & 3.49 -- 7.7 & 3.80 \\ 
& 2004.4 & 2.16 -- 2.57 & 3.51 -- 4.01 & 10.32  & & 2004.4 & 2.17 -- 2.83 & 3.64 -- 4.58 & 5.56 \\ 
& 2007.9 & 2.5 -- 4.1 & 2.58 -- 3.6 & 9.58  & & 2007.9 & 1.46 -- 3.58 & 2.45 -- 6.31 & 5.57 \\ 
& 2009.8 & 2.7 -- 3.54 & 2.95 -- 3.66 & 9.95  & & 2009.8 & 1.15 -- 4.0 & 2.35 -- 5.15 & 6.73 \\ \hline 
\multirow{4}{*}{R02\tablenotemark{*}} & 2000.2 & 2.02 -- 2.5 & 3.09 -- 4.0 & 10.24  &\multirow{4}{*}{R13} & 2000.2 & 1.73 -- 3.37 & 3.88 -- 7.04 & 3.03 \\ 
& 2004.4 & 2.75 -- 3.89 & 3.67 -- 4.59 & 5.66  & & 2004.4 & 1.45 -- 2.06 & 6.11 -- 9.9 & 3.17 \\ 
& 2007.9 & 3.33 -- 6.19 & 3.5 -- 4.58 & 7.54  & & 2007.9 & 1.39 -- 2.69 & 4.1 -- 10.7 & 3.04 \\ 
& 2009.8 & 3.37 -- 5.87 & 3.53 -- 4.64 & 9.65  & & 2009.8 & 1.11 -- 1.93 & 7.05 -- 22.35 & 4.24 \\ \hline 
\multirow{4}{*}{R03} & 2000.2 & 1.55 -- 1.7 & 5.16 -- 5.67 & 32.10  &\multirow{4}{*}{R14} & 2000.2 & 1.52 -- 2.24 & 3.45 -- 4.78 & 3.93 \\ 
& 2004.4 & 2.19 -- 2.5 & 4.66 -- 5.2 & 10.00  & & 2004.4 & 1.95 -- 2.31 & 4.26 -- 4.99 & 5.15 \\ 
& 2007.9 & 1.6 -- 1.92 & 5.25 -- 6.42 & 21.29  & & 2007.9 & 1.68 -- 2.54 & 3.4 -- 5.0 & 4.92 \\ 
& 2009.8 & 1.85 -- 2.4 & 4.51 -- 5.82 & 8.31  & & 2009.8 & 1.57 -- 3.15 & 2.39 -- 5.2 & 2.23 \\ \hline 
\multirow{4}{*}{R04} & 2000.2 & 1.7 -- 1.94 & 4.03 -- 4.71 & 30.65  &\multirow{4}{*}{R15} & 2000.2 & 2.07 -- 3.88 & 2.9 -- 4.1 & 2.08 \\ 
& 2004.4 & 1.8 -- 1.95 & 4.26 -- 4.67 & 34.09  & & 2004.4 & 2.02 -- 3.5 & 3.54 -- 4.72 & 2.40 \\ 
& 2007.9 & 1.73 -- 2.9 & 3.6 -- 5.6 & 14.45  & & 2007.9 & 1.24 -- 1.98 & 5.23 -- 9.25 & 5.68 \\ 
& 2009.8 & 1.75 -- 2.61 & 3.87 -- 5.57 & 10.96  & & 2009.8 & 0.97 -- 3.0 & 3.7 -- 9.3 & 2.58 \\ \hline 
\multirow{4}{*}{R05\tablenotemark{*}} & 2000.2 & 2.4 -- 2.9 & 3.17 -- 3.76 & 15.91  &\multirow{4}{*}{R16} & 2000.2 & 1.32 -- 1.5 & 18.0 -- 18.91 & 5.87 \\ 
& 2004.4 & 2.69 -- 3.36 & 4.14 -- 4.82 & 5.73  & & 2004.4 & 1.39 -- 1.6 & 15.2 -- 20.0 & 5.29 \\ 
& 2007.9 & 2.13 -- 3.65 & 4.54 -- 6.78 & 7.13  & & 2007.9 & 1.7 -- 2.63 & 8.07 -- 11.2 & 3.37 \\ 
& 2009.8 & 2.98 -- 3.69 & 4.22 -- 4.9 & 15.26  & & 2009.8 & 1.31 -- 1.94 & 10.52 -- 23.25 & 4.69 \\ \hline 
\multirow{4}{*}{R06} & 2000.2 & 1.25 -- 1.78 & 3.66 -- 5.4 & 6.87  &\multirow{4}{*}{R17} & 2000.2 & 2.0 -- 4.4 & 8.65 -- 15.0 & 1.08 \\ 
& 2004.4 & 1.89 -- 2.56 & 2.97 -- 3.81 & 3.09  & & 2004.4 & 2.71 -- 3.99 & 9.6 -- 13.37 & 1.07 \\ 
& 2007.9 & 2.99 -- 6.8 & 1.95 -- 2.93 & 2.10  & & 2007.9 & 1.8 -- 5.0 & 9.3 -- 27.15 & 1.21 \\ 
& 2009.8 & 1.52 -- 2.96 & 3.14 -- 5.65 & 3.57  & & 2009.8 & 1.92 -- 4.5 & 9.46 -- 18.75 & 1.18 \\ \hline 
\multirow{4}{*}{R07} & 2000.2 & 2.03 -- 2.4 & 2.83 -- 3.38 & 2.54  &\multirow{4}{*}{R18\tablenotemark{*}} & 2000.2 & 0.55 -- 0.62 & 23.12 -- 43.08 & 27.03 \\ 
& 2004.4 & 2.15 -- 4.4 & 2.27 -- 3.3 & 3.43  & & 2004.4 & 0.89 -- 1.1 & 8.07 -- 11.6 & 9.47 \\ 
& 2007.9 & 1.94 -- 2.64 & 2.65 -- 3.36 & 5.67  & & 2007.9 & 0.5 -- 3.0 & 3.82 -- 8.48 & 4.53 \\ 
& 2009.8 & 2.14 -- 3.95 & 2.52 -- 3.76 & 4.11  & & 2009.8 & 1.75 -- 3.0 & 5.5 -- 8.0 & 6.78 \\ \hline 
\multirow{4}{*}{R08} & 2000.2 & 3.89 -- 4.56 & 2.41 -- 2.65 & 1.42  &\multirow{4}{*}{R19} & 2000.2 & 1.53 -- 2.53 & 5.04 -- 8.2 & 3.35 \\ 
& 2004.4 & 1.6 -- 2.19 & 3.57 -- 4.59 & 2.66  & & 2004.4 & 1.35 -- 1.9 & 6.24 -- 9.9 & 3.77 \\ 
& 2007.9 & 1.11 -- 4.0 & 2.5 -- 6.1 & 1.88  & & 2007.9 & 1.34 -- 2.4 & 4.71 -- 11.15 & 3.41 \\ 
& 2009.8 & 1.1 -- 3.5 & 2.24 -- 3.78 & 4.37  & & 2009.8 & 1.1 -- 2.61 & 5.0 -- 32.4 & 4.52 \\ \hline 
\multirow{4}{*}{R09} & 2000.2 & 2.0 -- 3.08 & 12.51 -- 25.0 & 11.13  &\multirow{4}{*}{R20\tablenotemark{*}} & 2000.2 & 2.48 -- 2.96 & 2.92 -- 3.46 & 9.19 \\ 
& 2004.4 & 1.9 -- 2.9 & 14.0 -- 26.0 & 5.38  & & 2004.4 & 2.49 -- 2.85 & 3.22 -- 3.57 & 13.07 \\ 
& 2007.9 & 1.41 -- 2.48 & 16.0 -- 58.0 & 7.37  & & 2007.9 & 3.09 -- 3.75 & 2.73 -- 3.16 & 16.30 \\ 
& 2009.8 & 1.48 -- 2.53 & 15.92 -- 46.0 & 4.49  & & 2009.8 & 2.91 -- 3.8 & 3.13 -- 3.79 & 12.25 \\ \hline 
\multirow{4}{*}{R10} & 2000.2 & 0.89 -- 1.75 & 2.69 -- 5.5 & 2.04  &\multirow{4}{*}{R21\tablenotemark{*}} & 2000.2 & 0.76 -- 1.28 & 1.0 -- 5.5 & 0.66 \\ 
& 2004.4 & 0.79 -- 1.54 & 3.39 -- 9.2 & 2.05  & & 2004.4 & 1.51 -- 2.5 & 1.2 -- 2.29 & 0.52 \\ 
& 2007.9 & 0.94 -- 1.95 & 3.02 -- 5.4 & 2.58  & & 2007.9 & 2.37 -- 2.71 & 2.0 -- 3.11 & 0.80 \\ 
& 2009.8 & 1.18 -- 1.96 & 2.78 -- 8.2 & 1.41  & & 2009.8 & 3.31 -- 3.6 & 1.76 -- 2.17 & 1.26 \\ \hline 
\multirow{4}{*}{R11} & 2000.2 & 1.4 -- 1.89 & 5.31 -- 8.44 & 13.38  & & & & & \\ 
& 2004.4 & 1.81 -- 2.34 & 5.11 -- 6.59 & 7.72  & & & & & \\ 
& 2007.9 & 1.28 -- 2.12 & 5.06 -- 11.59 & 15.68  & & & & & \\ 
& 2009.8 & 1.77 -- 2.64 & 4.33 -- 6.71 & 15.08  & & & & &  
\enddata
\tablenotetext{*}{These knots had significantly non-overlapping confidence contours, and we consider them to be evolving.}
\tablenotetext{$\dagger$}{Norm values were not included in the confidence contour parameter estimation, so no uncertainties can be provided. For the sake of estimation, a limited study yielded one-dimensional confidence intervals ranging between $\pm 3\%$ for brighter knots and $\pm 50\%$ for dimmer ones.}
\end{deluxetable}
\clearpage

\begin{deluxetable}{cccccccc}
\tablewidth{0pt}
\tablecaption{The point estimates for the parameters of the ACIS analysis \\that were tied across all epochs\tablenotemark{a}.
\label{tab:acistied}}
\tablehead{ Region & $N_H  \hspace{3pt} \tablenotemark{b}$ & Mg & Si & S & Ar & Ca & Fe, Ni \\
         		 & ($10^{22} \mbox{ cm}^{-2}$)   & (solar) & (solar) & (solar) & (solar) & (solar) & (solar) }
\startdata
R01 & 1.52 & 0.32 & 3.97 & 3.36 & 3.21 & 2.20 & 0.05 \\  
R02\tablenotemark{*} & 1.63 & 0.12 & 5.04 & 4.41 & 4.26 & 3.87 & 0.00 \\  
R03 & 1.22 & 0.23 & 2.84 & 2.11 & 2.18 & 1.99 & 0.27 \\  
R04 & 1.23 & 0.34 & 2.67 & 2.06 & 2.84 & 1.42 & 0.27 \\  
R05\tablenotemark{*} & 1.23 & 0.11 & 2.93 & 2.13 & 2.21 & 3.15 & 0.16 \\  
R06 & 1.08 & 0.22 & 4.08 & 3.44 & 3.94 & 4.48 & 0.12 \\  
R07 & 1.31 & 0.00 & 6.52 & 7.51 & 7.47 & 12.20 & 0.48 \\  
R08 & 1.41 & 5.16 & 4.25 & 5.68 & 7.17 & 9.72 & 1.02 \\  
R09 & 1.53 & 0.58 & 4.88 & 3.92 & 3.54 & 2.47 & 0.79 \\  
R10 & 1.45 & 0.47 & 5.21 & 5.43 & 5.48 & 3.58 & 1.10 \\  
R11 & 1.63 & 0.45 & 3.13 & 2.58 & 2.09 & 2.74 & 0.60 \\  
R12 & 1.58 & 0.41 & 3.30 & 2.65 & 2.34 & 2.16 & 0.34 \\  
R13 & 1.49 & 0.48 & 3.95 & 4.11 & 4.51 & 3.97 & 0.64 \\  
R14 & 1.41 & 0.20 & 2.72 & 2.38 & 3.37 & 2.77 & 0.82 \\  
R15 & 1.12 & 0.18 & 4.60 & 4.09 & 4.65 & 3.72 & 0.27 \\  
R16 & 0.91 & 0.41 & 2.74 & 2.92 & 2.32 & 1.47 & 0.74 \\  
R17 & 0.93 & 1.00 & 6.12 & 5.58 & 4.95 & 2.86 & 4.70 \\  
R18\tablenotemark{*} & 1.42 & 0.31 & 1.44 & 1.62 & 1.63 & 1.22 & 0.00 \\  
R19 & 1.38 & 0.27 & 3.22 & 4.12 & 4.29 & 5.30 & 1.11 \\  
R20\tablenotemark{*} & 1.47 & 0.31 & 4.65 & 4.12 & 4.02 & 3.25 & 0.07 \\  
R21\tablenotemark{*} & 1.45 & 0.41 & 5.47 & 5.13 & 4.09 & 0.71 & 0.32
\enddata
\tablenotetext{*}{These knots had significantly non-overlapping confidence contours, and we consider them to be evolving.}
\tablenotetext{a}{No confidence contours were made for these parameters, so any apparent trends should be treated with reservations. Moreover, the general Si-rich, Fe-poor trends are biased, as the HETG analysis requires regions with bright Si features. C, N, and O were fixed to 5. The abundances are relative to the solar values of citet{ag89}.}
\tablenotetext{b}{The hydrogen-equivalent column density values are consistent with previous analyses (e.g. \citet{HL12}).}
\end{deluxetable}
\clearpage

\section{Appendix B: HETG analysis details}
\label{sec:HETGan}
\begin{figure}
\centering
\subfigure[]{
\includegraphics[angle=0,width=.56\columnwidth]{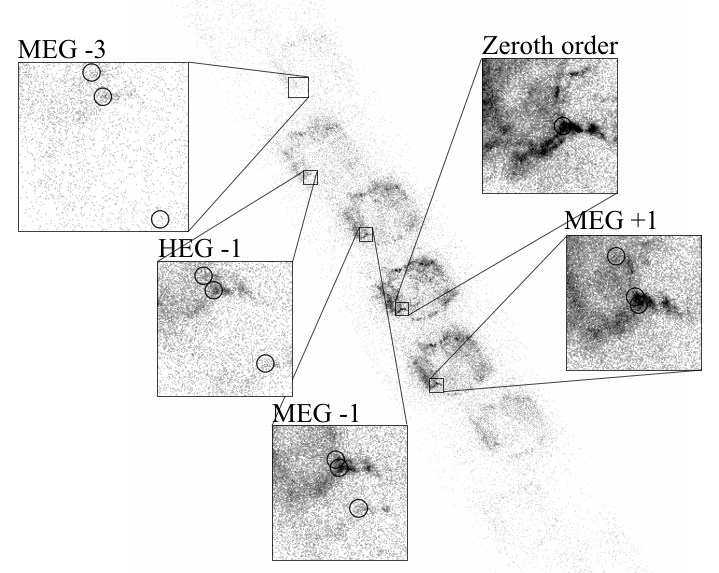}	
\label{fig:displocs}
}
\subfigure[]{
\includegraphics[angle=0,width=.40\columnwidth]{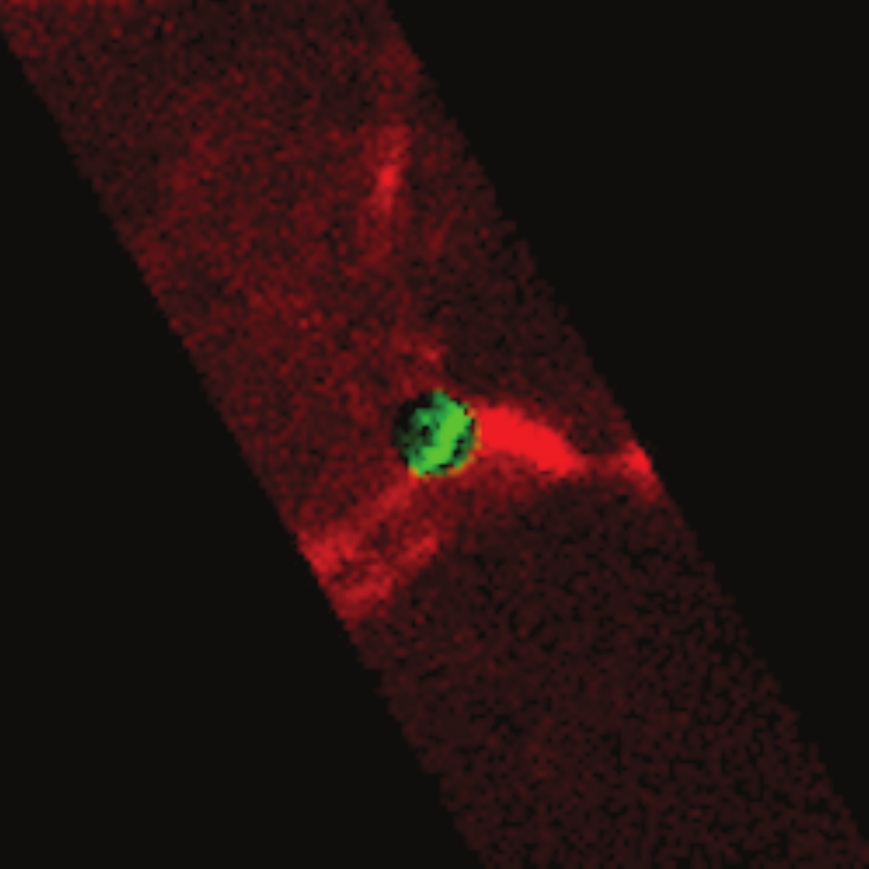}
\label{fig:zocomponents}
}
\caption{
The Event-2D software models and fits HETG dispersed data directly in two dimensions. (a) A knot (R1, in this case) can be found dispersed out to several orders. The circles show the expected dispersed locations of the H-like line and the He-like $r$ and $f$ lines for the best-fit velocity. (b) The zeroth-order spatial model for the same knot includes a 4-Gaussian spectral model (green, central region) and a \vnei\ component for the surroundings (red), based on the ACIS results. Including the surrounding component along the dispersion direction (upper-left to lower-right) better models the spatial-spectral overlap.\\}
\end{figure}

In general, extended sources cannot effectually be observed with slitless dispersive spectrometers such as the gratings on \chandra\ and \XMM.  However, for cases like the thermal knots of Cas~A, their small size and bright line emission allow line information to be extracted from the dispersed data\,\citep{Dewey02}.  On CCD-readout instruments, the dispersed images (Figure\,\ref{fig:displocs}) carry useful information not only in the direction of dispersion, but also in the cross-dispersion axis, as well as in energy, which can provide order sorting.

The previous analyses of \Laz\ adapted the usual 1D spectral fitting machinery of \pha, \arf, and \rmf\ files to the Cas~A knots by collapsing the dispersed events along the cross-dispersion direction. To achieve higher resolution, the features were bent along the cross-dispersion axis before creating that order's 1D spectrum. The companion \rmf\ for each order encoded the spectral broadening introduced by the spatial distribution of the sheared zeroth-order events.  A spectral model consisting of continuum plus several lines was fit to the \pha\ data to determine Doppler shifts and line ratios.

For the current (re-)analyses of the HETG datasets, we improve the analysis technique by modeling and fitting the 2D dispersed events directly.  The Event-2D software\,\footnote{Event-2D web page: http://space.mit.edu/home/dd/Event2D/}, briefly described in \citet{Dewey09}, is written in S-Lang\,\footnote{S-Lang web page: http://www.jedsoft.org/slang/} to provide an extension of the ISIS software. The software is an example of X-ray analysis that goes beyond the usual 1D fitting approach\,\citep{Noble08}. Event-2D removes the need to define a filament, allows different spectral models to be assigned to the knot and its surroundings, and, through its instrument simulation, utilizes a narrow order-sorting range to reduce background.

\subsubsection{Knot extractions}

Spectral extraction consists of generating the usual spectral products (\pha, \arf, and \rmf\ files) for a source observed with the HETG.   The standard extraction process also adds grating-specific data (columns) to the \evt\ file based on the location of the source center. These values include an event's assigned diffraction order and 2-dimensional grating coordinates, {\tt TG\_R} and {\tt TG\_D}. (Note that many events will not be part of an extraction for a given source location; these events are flagged by a diffraction order set to 99.) It is this \evt\ file that Event-2D uses as input data in place of the \pha\ file. Because of this, the standard HETG extraction steps are sufficient preparation for the Event-2D analyses.

The extractions for the Cas A knots were carried out with CIAO tools using standard HETG-appropriate steps. For convenience, scripts from TGCat\footnote{TGCat web page: http://tgcat.mit.edu/ } were used to simplify this process, yet still allow customization. The location for the extraction of each knot was manually set by comparing the regions of ACIS epoch 2004.4 with the HETG zeroth-order image to effectively select the same knot. We made some adjustments to the usual extraction parameters to improve the subsequent 2D analyses: (1) the cross-dispersion width was set to twice the knot diameter, (2) a large zeroth-order radius of 60 arcseconds was defined, (3) a narrow order-sorting range of $\pm$\,0.06 was set, and (4) \arf\ files were made for orders $m = \{0,\,\pm 1,\,\pm 2,\,\pm 3\}$\,. The large zeroth-order range was chosen to include (and hence model) off-knot spatial features that can produce spatial-spectral overlap with the knot. The narrow order sorting range still includes most of the Si-range flux while helping to reduce cross-talk from other spatial regions.

\subsubsection{2D spatial-spectral modeling}
\label{hetgmodel}

The spatial model of each knot is self-described using the zeroth-order events within a given radius of the extraction center. The events surrounding the knot region are used to define a second spatial component of the model -- see Figure~\ref{fig:zocomponents}.  Modeling this surrounding region is an important part of generating simulated dispersed events in the wavelength range of interest.

Spectral models are assigned to the two spatial components. The knot spectrum is described by 4 Gaussians corresponding to the He-like $r$, $i$ and $f$ lines and the H-like Ly\,$\alpha$ line.  The $f/i$ ratio is fixed at the low density value of 2.45 and the remaining free parameters are then the overall flux, the f/r ratio, the H/He ratio and a common Doppler shift. For the surroundings' spectrum we start with the best-fit ACIS (Epoch 2004.4) \vnei\ model and allow its $\tau$ to be adjusted. This can give a reasonable approximation of the surroundings, especially for the closest features that are most important. (Of course one could explicitly measure and use the spectrum of the surrounding region itself, perhaps even in multiple zones.  However, we have not seen a clear need for this level of model fidelity in the analyses so far.) Because we are not very sensitive to the continuum shape, the temperature is held fixed. Although it need not be, the velocity of the surroundings is fixed at the knot velocity; we have not seen a clear indication in the data for a need to change this.

\begin{figure}
\begin{center}
\includegraphics[angle=0,width=.5\columnwidth]{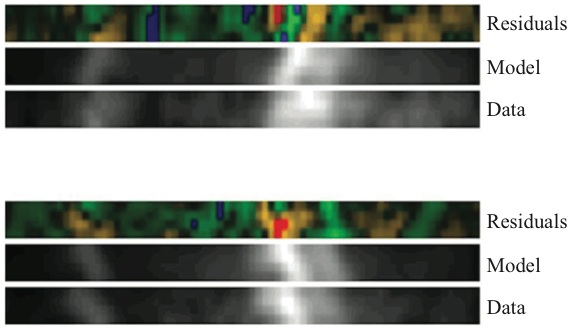} 
\end{center}
\caption{
The Event-2D software shows remarkable agreement between data and the model. Shown are the epoch 2010.3 R5 MEG $\pm1$ order 2D images.  The Si-line region is shown with the wavelength range going from shortward of the Si~XIV line at left to longward of the Si~XIII triplet at right.  Note how the apparent angle of the knot differs between orders.
\label{fig:2Ddatamodel}}
\end{figure}

The ``Source-3D'' routines (companion to Event-2D) are used to organize and access the spatial-spectral model described above. In particular Monte Carlo source events can be generated from the defined model and then passed through an approximate HETG instrument simulator in Event-2D.  In this way simulated model data are created that can be compared with the actual data (Figure\,\ref{fig:2Ddatamodel}). The  inclusion of order-sorting effects in the instrument model is notable; the effects play a large role in shaping the observed and modeled events.

\subsubsection{Parameter estimation}

Parameter estimation proceeds by forward-folding the spatial-spectral model to generate model events, which are then compared to the spectra. The data and model events are binned in defined 2D histograms, and the chi-squared statistic is calculated: $\chi^2 \,=\, \sum_i\,((D_i-M_i)/\sigma_i)^2$, where $\sigma_i=\sqrt{0.63\,M_i+0.37\,D_i}$.  This modified definition of $\sigma_i$ reduces the bias in the statistic compared to either $\sigma_i=M_i$ or $\sigma_i=D_i$.

The randomized nature of the model generation forces us to use noise-tolerant fitting algorithms. A small change in a parameter value can be masked by the ``model noise'' of Monte Carlo samples (even though the model counts are over-simulated by a factor of 10).  For instance, around the best-fit value of parameter $p$, we have that $\Delta \chi^2 \approx a\,\Delta p\,^2 + \sigma_{\chi^2}$ where $a$ is a constant depending on the particular data and model and the $\sigma_{\chi^2}$ term represents the noise in the model computation. Hence changes in $p$ are not ``noticeable'' above the model noise unless $\Delta p > \sqrt{\sigma_{\chi^2}/a}$, effectively setting a  minimum scale size for parameter $p$. Because of this, we use minimization schemes that include a minimum scale for noticeable changes in a parameter and that are tolerant of the model noise. For example, to determine a single parameter like the Doppler velocity, multiple evaluations of the statistic are performed at each step of 100 $\mbox{km s}^{-1}$ to estimate $\sigma_{\chi^2}$, which aids in determining the best parameter value. For multi-parameter fitting, a noise-tolerant conjugate-gradient (C-G) scheme can be constructed to minimize a parameter in 1D and include the noise level in its convergence diagnostics.

Another approach to dealing with noisy models is to use a MCMC method: the noise of the model generation is only a slight addition to the larger variation expected as the parameter space is randomly explored and so the technique is not sensitive to the model noise.

To analyze these X-ray knots, we combine the two methods. The C-G fitting coarsely sets the surrounding region norm and Tau values, fixing the Tau value. A 1D scan in velocity space is then performed and the velocity is fixed.  A C-G scan is then done to fit the two line ratios and the ratio of flux from the surrounding region to that of the knot. MCMC exploration is carried out for these 3 parameters starting from the C-G best-fit values; in this way there is generally little settling time for the MCMC.  The spread in the MCMC draws provides the error bars for the parameters. 

As mentioned above,  the modeling is sensitive to the CCD gain calibration through the narrow order-sorting range.  After a knot's analysis is complete, the simulated and real CCD energy distributions are compared for the bright He-like Si line in each order.  Where these differ by more than 0.8\%, a custom gain correction is included for that epoch-knot-order (applied when the data are read in) and the knot is re-analyzed.  

The measured values for a knot are its Doppler velocity, the two line ratios ($f/r$ and He-like/H-like), and the overall knot flux normalization; these are given in Table~\ref{tab:HETGresults} for both epochs of HETG observations.  Uncertainties are given for all but the flux, which has a statistical uncertainty generally less than 2\% and so will be dominated by systematic errors, e.g., calibration of the effective area at a knot's specific location.

\section{Appendix C: ACIS analysis details}
\label{sec:ACIS-anal}
In this appendix we provide the details of the data reduction and analysis of the ACIS data. The observation IDs and exposure times are listed in Table~\ref{tab:ACISobs}, Section~\ref{sec:obs}. The ACIS analysis used four archival datasets from early 2000, mid-2004, late 2007, and late 2009. All four of these observations were taken on the back illuminated ACIS-S3 chip in GRADED data mode.  They were targeted at a right ascension of $23^{\mbox{h}} 23^{\mbox{m}} 26^{\mbox{s}}.70$ and a declination of $+58^{\circ} 49^{\prime} 03^{\prime\prime}.00$.

\subsubsection{Spectral definition of the knots}
\label{sec:regsearch}

We assumed these small, bright knots are individual clumps of similar material, so we therefore defined their boundaries spectrally. Once defined, all of the enclosed material shares the same spectral features. Arbitrarily bounding these regions by hand, using only brightness information and an analyst's intuition, could unintentionally include non-knot material of similar surface brightness, thereby skewing results of the analysis. We found this to be the case, so developed a more systematic method than hand-selection. 

\begin{figure}
\begin{center}
\includegraphics[angle=0,width=\columnwidth]{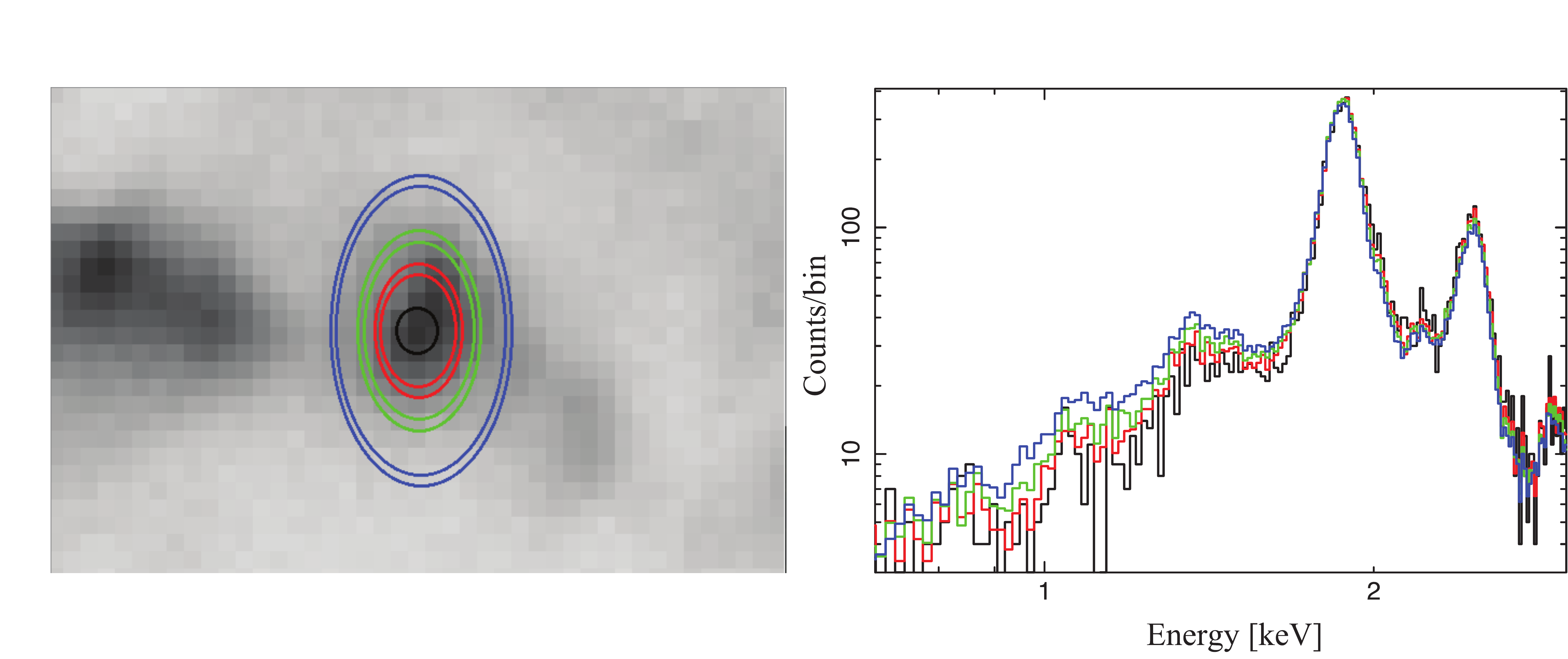} \\
\end{center}
\caption{The knots are defined spectrally. Left: Anulii and spectra used for the knot region definition for knot R02. The two same-color ovals define an annulus, and the spectra from inside this annulus has been extracted and compared to the central black oval's  spectrum. Right: normalized spectra from the central part of R02 (black) and three annuli. A soft-energy excess can be seen in the spectral shape of the outer annuli. These spectra were used to determine the boundary of the knot ensuring spectral homogeneity within the defined knot.\\
\label{fig:regsearch}}
\end{figure}

Our region-determining algorithm finds the radius at which nearby material starts to contaminate the knot. For each knot we select an inner region consisting of the brightest central pixels -- typically 4 to 6 in number; this region defines the spectrum of the knot. We then compare this spectrum to the spectra of annuli of increasing radius (Figure~\ref{fig:regsearch}), evaluating the Poisson likelihood of the annulus data, given the central region's spectrum as the model. The knot boundary is defined at the annulus where the likelihood drops sharply, or where the normalized counts per unit area drop abruptly. In the absence of sharp changes, we visually compared the spectra to delineate the boundary. 

\subsubsection{The use of close-in backgrounds}

\begin{figure}
\begin{center}
\includegraphics[angle=0,width=.5\columnwidth]{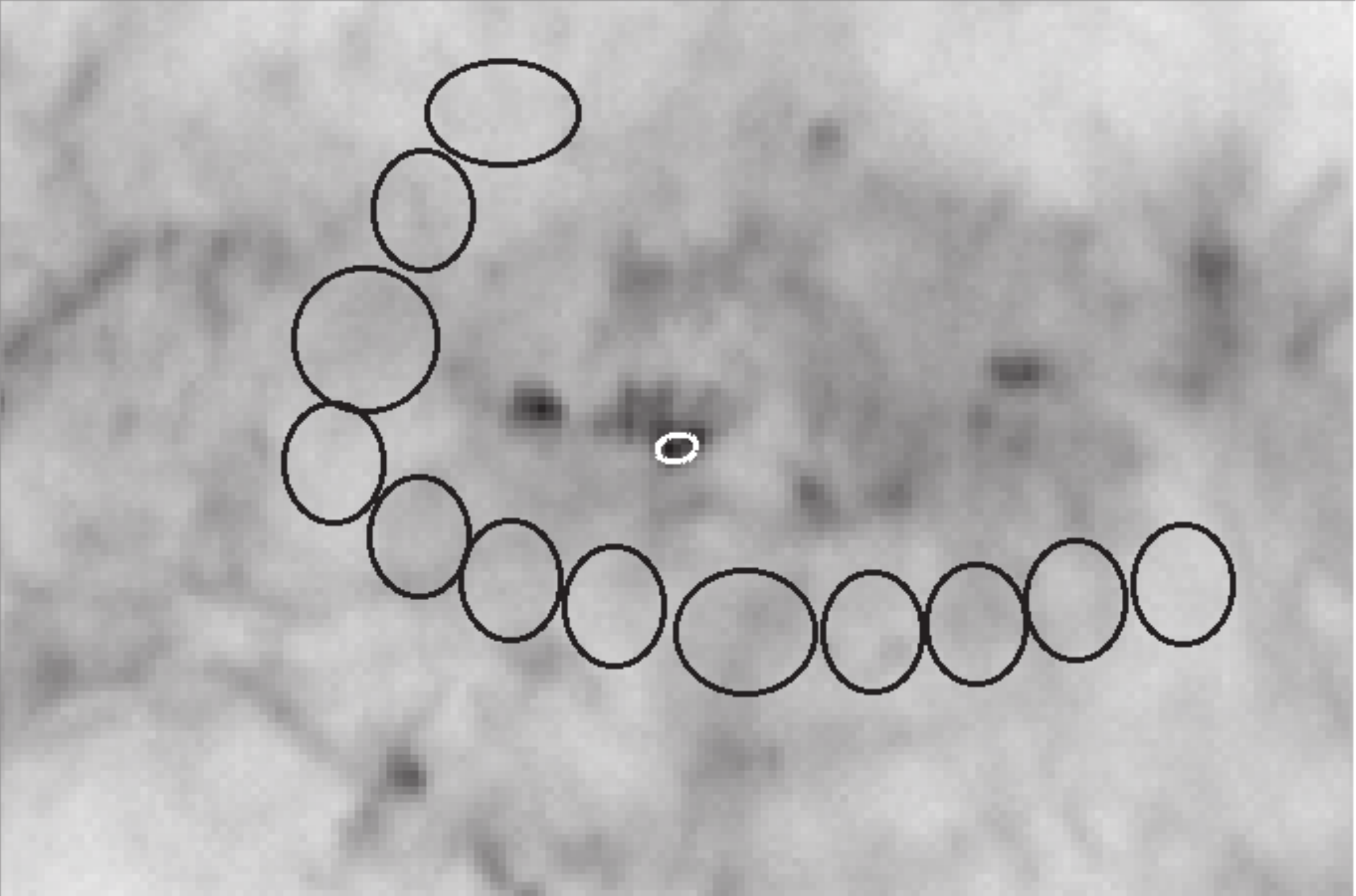} \\		
\end{center}
\caption{We assume that the low-brightness surrounding material fills the front or rear of the knot, at about the same level as in our background region. The background regions are shown in black with the R05 region shown in white. By taking many regions around the knot, we effectively average out the local small-scale variations to approximate the actual backgrounds and foregrounds in the spectral extraction ``core.'' \\
\label{fig:knotbkgd}}
\end{figure}

The bright X-ray knots are expanding amidst a diffuse web of emitting material. We must therefore carefully consider how to define the ``background'' (which could very well be the foreground) for spectral modeling, as the knot is seen \emph{through} this web. We systematically examined the characteristics of the surrounding material to determine the most faithful way to define the background for each knot. 

We decided to assign multiple diffuse, nearby regions as the background for each knot. Figure~\ref{fig:knotbkgd} shows a typical background assignment. These diffuse regions were characterized by low surface brightness (over the \chandra\ passband) and variations over larger angular scales. We typically chose several regions for each knot for two reasons. First, these regions were more spectrally consistent (and therefore yielded similar plasma parameters for the knot model) than background regions that contained brighter features varying on smaller angular scales. Second, since this diffuse material varied on longer scales, it was more probable that it contaminated the knot spectrum. Choosing multiple regions around the knot effectively interpolated the spectrum to the position of the knot. The regions remained the same across epochs, save for slight nudging required because of the ejecta expansion. 

The background regions were chosen near to the knots but not directly adjacent. We considered using the discarded non-knot annuli from the region-determining algorithm. However, the material in these adjacent rings showed spectral features not present in the knot on order of the same surface brightness, suggesting different compositions. 

In general, we find that the diffuse material we select as backgrounds tends to be cooler and to move more slowly than the ejecta knots.  Including background spectra in the model generally results in higher derived plasma temperatures and Doppler velocities for the knots. 

\subsection{Data Extraction}

The development of this algorithm yielded a unique image extraction technique.  For an extended source like Cas~A, the \specextract\ \ciao\ script is typically used. The area-weighted \arf\ (\warf) created by \specextract\ does not include the effect of angular dither, since the region area is assumed to be large compared to the dither.  The annuli usually have a size of 6-15 pixels, large enough to warrant a \warf\ and \wrmf, yet small enough to be affected by the dither of the instrument~\citetext{David P. Huenemoerder, private communication, 2010}. To create, then, a \warf\ that includes dither, we first make the \texttt{pi} file and \wrmf\ with \specextract.  Running the \texttt{sky2tdet} function outputs a \texttt{wmap} file describing the dither, which can then be used a second time in \texttt{mkwarf}, rewriting the \warf. The tweaked calibration product is now appropriate to our peculiarly sized regions.   
This extraction was done using CIAO version 4.3, with CALDB version 3.2.2.

\subsection{Spectral Fitting}

We applied a non-equilibrium ionization model with variable abundances (\vnei\ version 2.0) to the data, also accounting for interstellar absorption (\phabs) and pileup \citep{Davis:2001js}.  We chose not to employ the similar \texttt{vpshock} model due to the additional free parameter of the lower $\tau$ value. Subsequent investigations on several knots yielded maximum likelihood \texttt{vpshock} values of $\tau$ and $kT$ within 10\% of those for the \vnei\ model, where we compared the upper $\tau$ for \texttt{vpshock} to twice that of the \vnei\ $\tau$. Therefore, we do not think our choice of model unduly biased the results.

For each knot, we expect several physical parameters to stay constant across the decade of observations: the equivalent neutral hydrogen column density and the elemental abundances.  To this end we performed the fits to the four epochs of ACIS data for each knot simultaneously, while tying $n_{H}$ and the abundances from the 2004.4, 2007.9 and 2009.8 datasets to that of the 2000.2 dataset.  Values for C, N and O abundances were frozen to 5 times the solar value. (As discussed in Section~\ref{sec:vneidegen}, the actual values of the abundances used can be degenerate with the norm of the continuum fit. We thus chose a value of 5 to account for the expected C, N, and O dominance in the continuum. \citealt{Vink:1996wp} argues for O, while \citealt{Dewey:2007ue} argues for H/He, or C/N.) Fe and Ni abundances were tied together. 
 
For the \texttt{pileup} component of the model, all parameters but $n$---the number of regions where pileup occurs---are fixed across epochs: $\bar{g_0} = 1$, $\alpha = 0.5$ and $f = 1$. The $\bar{g_0} $ and $\alpha$ values are simply fiducial estimates for grade-migration processes. The parameter $f$ was set to 1 because all of the knots are larger than Chandra's PSF. Letting $n$ alone vary captured the effects of pileup well, reducing the $\chi^2/\mbox{d.o.f.}$

The redshift parameter in the model was treated only as a calibration gain factor, and it was allowed to differ from epoch to epoch. Redshifts were point-estimated at the beginning of the fitting process, then frozen throughout the confidence contour generation. This initial point estimate was fit with the He-like $r$, $i$, $f$,  and H-like Ly\,$\alpha$ Si lines folded through the ACIS \rmf.  By fixing the redshifts to these maximum likelihood values, the confidence contour computation time was greatly reduced. (We attribute the added computational complexity to the discontinuous $\chi^2$ space along the redshift axis, a persistent feature we found with this model.)
 
The combined four-epoch fit for each knot involved 88 parameters, with 25 of these free. With such a complicated parameter space, it was easy for the fit to fall into---and get hung up in---local minima.  In order to better find the global minimum for each fit, we used the native, parallelized ISIS function \confloop.  This function performs confidence limit searches on each parameter individually about the current best fit value, often climbing out of a local minimum to find a better fit. 

\subsection{Generating the confidence contours}

Once the spectral fits have found the maximum likelihood point for each epoch of each knot, we proceed to generate confidence contours for our fit parameters. A grid is laid down around this best estimate parameter vector and the $S$ statistic (defined below) is minimized at each point to map out the confidence levels. The confidence contours are made with \texttt{conf\_map\_counts} in ISIS. 

The results of these fits are shown in Appendix~\ref{sec:result-app}, Figures~\ref{fig:acisresults1} -- \ref{fig:acisresults3} and Tables~\ref{tab:ACISresults} and \ref{tab:acistied}. 

\begin{figure}
\begin{center}
\includegraphics[angle=0,width=2in]{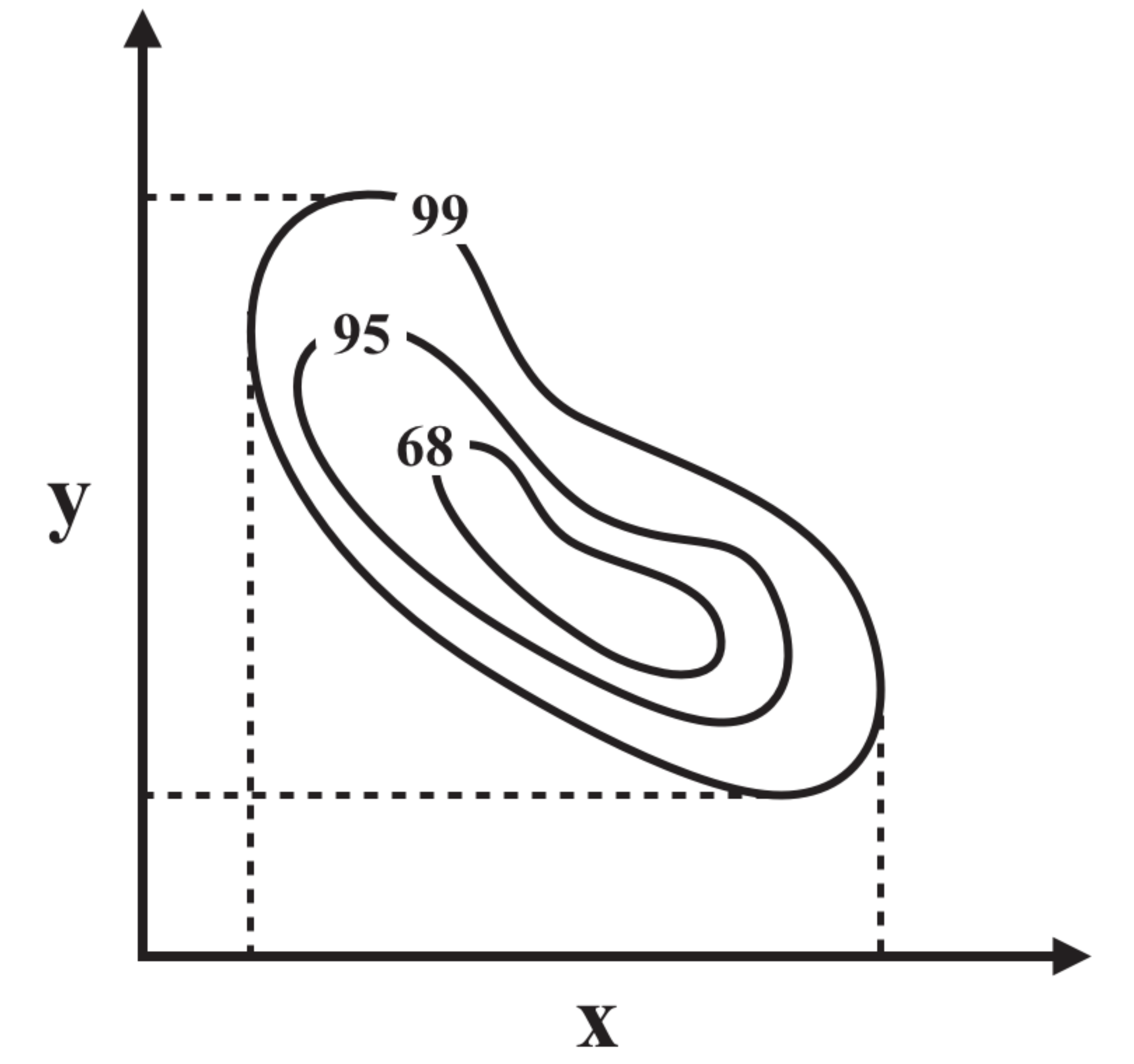} \\
\end{center}
\caption{We report our uncertainties for the ACIS analysis as bounding edges of the confidence contours. Most of the knots exhibit correlated ``bananas'' in $kT$--$\tau$ space, so the summary is sufficient. \\
\label{fig:ccedges}}
\end{figure}

The reported parameter ranges represent confidence levels. That is, if the analysis could be repeated, the true parameters would lie within the 90\% contours for 90\% of the re-analyses. Moreover, we only report the bounding edges (see Figure \ref{fig:ccedges}), even though the confidence contours are never rectangular. We do this for conciseness and because the contours all have the same anti-correlated ``banana'' shape seen in Figure~\ref{fig:acisresults1}.

\subsection{Assumptions and choice of fit statistic}

We make several assumptions to warrant our use of confidence contours for parameter estimation. 
We assume the counts in each energy bin are independent from each other and that they are Poisson-distributed, with each energy bin $E_{i}$ having a mean $\lambda_i$ given by the model. We group counts into energy bins of at least 30 counts, so that we may approximate the count distribution in each bin as normal, with mean $\lambda_i$ and standard deviation $\sqrt{\lambda_i}$. (Our inclusion of pileup both undermines and supports this assumption. We admit that a count in one bin may originate from two in another, yet since we account for pileup effects, the bins of the model spectra will have less correlation.)

This large minimum number of counts per energy bin allows us to employ the $S$ statistic, popularized by Cash for nonlinear X-ray models. (We did not use Cash's more appropriate $C$ statistic solely because of the extra computation time within the ISIS framework.) The $S$ statistic is merely --2 times the log likelihood:
\[S = \sum_{i = 0}^{N-1} \frac{(d_i - \lambda_i)^2}{\lambda_i},\]
where $d_i$ are the data counts in each of the $N$ bins and, as above, $\lambda_i$ are the predicted model counts. We are interested in only a subspace of the whole parameter space, primarily the $kT$s and $\tau$s over the four epochs. Cash's result says that the  difference between the minimized values of $S$ over this $q$-dimensional subspace and the overall best fit $S$ follows a $\chi^2_q$ distribution. Thus, with probability $1-\alpha$, the true values of the parameters reside within the contour(s) where this difference equals $\chi^2_q(\alpha)$.

As a final approximation, during confidence contour generation we assume that the epochs are independent, despite explicitly tying $N_H$ and elemental abundances between epochs. That is, for one epoch of one knot, we grid up $kT$ and $\tau$ values and minimize $S$ at those points, with $N_H$, the abundances, and the $kT$s and $\tau$s from other epochs free to vary. We do this merely for computational tractability: under this approximation we only have to minimize $S$ on $4 \times N^2$ grid points, instead of $N^{2 \times 4}$. The maximum likelihood parameters for each epoch differed little from the global best parameters, suggesting this assumption is sound. 

\subsection {Electron density derivation}

To calculate $n_e$, we use a combination of model parameters and inferred values. Parameter estimation with the \vnei\ model yields the norm, $X_{norm}$, and the abundances, $X_A(Z)$. We estimate the knot volume, $V$, as an oblate spheroid, with axes taken from our region-finding algorithm. We set the distance to Cas A to $d = 3.4 \mbox{ kpc}$. We approximate the number of electrons stripped with a functional form:
\[Q(Z, T) = 0.41 Z (\log(T(28/Z)^2) - 5.2) .\]
(The function is constrained to the interval [0, $Z$].) Finally, the solar abundances, $A(Z)$, are drawn from \citet{ag89}. 

Using the conventional definition of $X_{norm}$, we can find the electron density.

\begin{eqnarray*}
n_e &=& \sqrt{ (n_e n_H)\left(\frac{n_e}{n_H}\right) } \\
&=& \sqrt{ \frac{4\pi d^2 10^{14}X_{norm} X_A(Z=1)}{V} \left(\frac{n_e}{n_H}\right) } \\
&=& \sqrt{ \frac{4\pi d^2 10^{14}X_{norm} X_A(Z=1)}{V} \frac{\sum{Q(Z, T)X_A(Z)A(Z)}}{X_A(Z=1)A(Z=1)} } 
\end{eqnarray*}

The factor of $X_A(Z=1)$, called $X_H$ in the main text, properly scales the norm, allowing us to set the hydrogen abundance in \vnei\ to something other than 1.

Finally, if we deem than a \vnei-modeled plasma component does not take up the full volume of the knot, then $V \rightarrow fV$, where $f$ is the fill fraction. 

\end{document}